\documentclass{aa}  

\usepackage{graphicx}
\usepackage{txfonts}
\begin{document}

   \title{Long-term photometry of the eclipsing dwarf nova 
                        V893~Scorpii\thanks{Based 
                        on observations taken at the
                        Observat\'orio do Pico dos Dias / LNA}  }

   \subtitle{Orbital period, oscillations, and a possible giant planet}

   \author{Albert Bruch}

   \institute{Laborat\'orio Nacional de Astrof\'{\i}sica,
              Rua Estados Unidos 154, CEP 37504-364, Itajub\'a,
              Brazil\\
              \email{albert@lna.br}
             }

   \date{Received <date> / Accepted <date> }

  \abstract{The cataclysmic variable V893~Sco is an eclipsing dwarf nova 
which, apart 
from outbursts with comparatively low amplitudes, exhibits a particularly 
strong variability during quiescence on timescales of days to seconds.}
{The present study aims to update the outdated orbital ephemerides 
published previously, to investigate deviations from linear ephemerides, and to 
characterize non-random brightness variations in a range of timescales.}
{Light curves of V893~Sco were observed on 39 nights, spanning a total 
time base of about 14 years. They contain 114 eclipses which were used to
significantly improve the precision of the orbital period and to study
long-term variations of the time of revolution. Oscillations and similar
brightness variations were studied with Fourier techniques in the
individual light curves.}
{The orbital period exhibits long-term variations with a cycle time of 10.2
years. They can be interpreted as a light travel time effect caused by the 
presence of a giant planet with approximately 9.5 Jupiter masses in a 4.5 AU 
orbit around V893~Sco. On some nights transient semi-periodic 
variations on timescales of several minutes can be seen which may be 
identified as quasi-periodic oscillations. However, it is difficult to
distinguish
whether they are caused by real physical mechanisms or if they are the effect 
of an accidental superposition of unrelated flickering flares. Simulations to
investigate this question are presented.}{} 


   \maketitle
%

\section{Introduction}
\label{Introduction}

Cataclysmic variables (CVs) are well known to be short period 
interacting binary systems where a Roche-lobe filling
star, the secondary, transfers
matter via an accretion disk to a white dwarf primary.
The structure of CVs can best be studied in systems
where the secondary, which is faint and contributes in most
cases only negligibly to the optical light, eclipses periodically 
the bright accretion disk and the white dwarf. 
In systems like these it is possible not only
to derive the orbital period easily and accurately, but also to
set more stringent limits on many fundamental system parameters.
Moreover, eclipses frequently can be used as a tool to study 
structural details of the binary system and its components.

\object{V893 Sco} is a member of the dwarf nova subclass of CVs. 
It was identified as a variable star and classified by 
Satyvoldiev (\cite{Satyvoldiev}) but got lost thereafter. 
Only much later, in 1998, Kato et al.\ (\cite{Kato}) re-identified
V893~Sco. Soon thereafter, Bruch et al.\ (2000; hereafter referred to
as Paper~I) published a photometric study, reporting the discovery of
eclipses and deriving an orbital period of $1^{\rm h}\, 49^{\rm m}\, 23^{\rm s}$
which makes the system a member of short-period CVs located below the
famous gap in the orbital period distribution of cataclysmic variables.
Moreover, they found strong cycle-to-cycle variations concerning the mean 
magnitude, the strength of the orbital hump, the presence of an intermediate 
hump, and the amplitude and minimum depth of the eclipses. Time-resolved
optical spectroscopic studies of V893~Sco were published by Matsumoto
et al.\ (\cite{Matsumoto}) and Mason et al.\ (\cite{Mason}) 
and show that the system -- in spite
of exhibiting some peculiarities -- is spectroscopically similar to many
other CVs of the same kind. 
Thorstensen (\cite{Thorstensen}) measured trigonometrically a distance of 
$153^{+68}_{-35}\, {\rm pc}$. Mukai et al.\ (\cite{Mukai}) observed 
V893~Sco in x-rays with the {\em Suzaku} 
satellite and found partial eclipses. Warner et al.\ (\cite{Warner03}) report on
the detection of quasi-periodic oscillations (QPOs) [which have also been seen 
by Bruch et al.\ (2000)] and dwarf nova oscillations (DNOs), and Pretorius
et al.\ (\cite{Pretorius}) claim to have observed an instant with DNO activity.

In continuation of the observations discussed in \cite{PaperI},
V893~Sco was observed regularly over the past 14 years. In this study I
present some of the results of this long-term effort.
In Sect.~\ref{Observations}, the reader is introduced to 
the observations. The photometric state during which they were taken and the
characteristics of the long-term light curve are briefly discussed 
in Sect.~\ref{Photometric state}.
The partially eclipsed x-ray source in V893~Sco (Mukai et al., 
\cite{Mukai}) is
expected to be located in the immediate vicinity of the white dwarf (e.g.\ a 
boundary layer). Therefore, the suspicion of Bruch et al.\ (\cite{PaperI}), 
based on the strong variability of the amplitude and minimum depth of the 
eclipses, that the
eclipsed body is the hot spot and that the centre of the accretion disk
and the white dwarf remain uneclipsed, is not tenable. This is the topic
of Sect.~\ref{Eclipse profile} which is dedicated to an analysis of the eclipse
profile. Mukai et al.\ (\cite{Mukai}) already noticed that the ephemerides 
derived in \cite{PaperI} did not well predict the eclipse times at the epoch
of their x-ray observations. Thus, an update is required. This is done
in Sect.~\ref{Ephemerides} using data collected over a vastly longer time base.
The new ephemerides 
reveal cyclic period variations which are investigated in detail and
interpreted in Sect.~\ref{The cyclic period variations}. The presence of a
third body with the mass of a giant planet can explain the observations. 
The detection of QPOs and DNOs by Bruch at al.\ (\cite{PaperI}), Warner
et al.\ (\cite{Warner03}) and Pretorius et al.\ (\cite{Pretorius}) 
justifies a frequency analysis of the light curves which is performed in 
Sect.~\ref{Frequency analysis}. 
Here, I also address the question of the
reliability of peaks in power spectra in the presence of strong flickering
activity. Finally, the results of this study are summarized in 
Sect.~\ref{Conclusions}.

\section{Observations}
\label{Observations}

All observations presented in this paper were obtained at the 0.6-m Zeiss
telescope of the 
Observat\'orio do Pico dos Dias, operated by the Laborat\'orio Nacional de
Astrof\'{\i}sica, Brazil. A complete summary of the observations
is given in Tab.~\ref{Journal of observations}. They include also the
high-speed light curves discussed already in \cite{PaperI}.

\begin{table}
\caption{Journal of observations.}
\label{Journal of observations}
\centering

\begin{tabular}{l@{\hspace{1ex}}c@{\hspace{1ex}}
c@{\hspace{1ex}}c@{\hspace{1ex}}c@{\hspace{1ex}}c@{\hspace{1ex}}}
\hline\hline

Date            & Julian    & Start & End   & Time & Number  \\
                & Date      & (UT)  & (UT)  & Res. & of      \\
                & (2450000+)&       &       & (s)  & Integr. \\
\hline
1999 May 09\tablefootmark{a}    & 1307      & 00:54 & 06:42 & 5    & 1463    \\
1999 May 24/25\tablefootmark{a} & 1323      & 23:35 & 07:43 & 3    & 8682    \\
1999 Jun 11/12\tablefootmark{a} & 1341      & 22:32 & 06:28 & 3    & 6435    \\
1999 Jun 21/22\tablefootmark{a} & 1351      & 22:39 & 06:12 & 3    & 5291    \\
2000 Apr 25    & 1659      & 01:37 & 03:37 & 4    & 1812    \\ [1ex]
2000 Apr 26    & 1660      & 01:45 & 03:44 & 5    & 1411    \\
2000 Apr 27    & 1661      & 01:31 & 03:37 & 5    & 1496    \\
2000 Apr 28    & 1662      & 01:20 & 03:25 & 5    & 1494    \\
2000 May 06    & 1670      & 00:13 & 02:48 & 5    & 1845    \\
2000 May 24    & 1688      & 00:03 & 08:17 & 5    & 5187    \\ [1ex]
2000 May 24/25 & 1689      & 23:14 & 08:08 & 5    & 6335    \\
2000 May 25/26 & 1690      & 22:34 & 06:54 & 5    & 4263    \\
2000 May 27    & 1691      & 00:34 & 07:57 & 5    & 5283    \\
2000 Jun 06/07 & 1702      & 21:50 & 07:18 & 5    & 6562    \\
2000 Jun 07/08 & 1703      & 21:29 & 07:18 & 5    & 6833    \\ [1ex]
2000 Jun 08/09 & 1704      & 21:35 & 07:18 & 5    & 6732    \\
2000 Jun 09/10 & 1705      & 23:59 & 07:13 & 5    & 5053    \\
2001 Mar 16    & 1984      & 05:00 & 07:37 & 5    & 1854    \\
2001 Apr 06    & 2005      & 02:01 & 02:48 & 5    & \phantom{0}314 \\
2001 Jun 29/30 & 2090      & 23:12 & 05:42 & 5    & 4420    \\ [1ex]
2002 May 15/16 & 2410      & 23:25 & 07:58 & 5    & 6120    \\
2002 Jun 24/25 & 2450      & 23:17 & 05:43 & 5    & 3911    \\
2003 May 09/10 & 2769      & 23:45 & 03:46 & 5    & 2852    \\
2003 Jun 20/21 & 2811      & 21:48 & 06:24 & 5    & 5563    \\
2004 Mar 17    & 3081      & 03:41 & 08:29 & 5    & 3443    \\ [1ex]
2004 Jun 22/23 & 3179      & 23:19 & 06:10 & 5    & 4648    \\
2004 Jul 13/14 & 3200      & 21:52 & 04:20 & 5    & 4534    \\
2004 Jul 14/15 & 3201      & 22:44 & 01:31 & 5    & 1522    \\
2005 May 07    & 3497      & 02:10 & 07:34 & 5    & 3260    \\
2005 Jun 17/18 & 3539      & 22:12 & 06:35 & 5    & 5288    \\ [1ex]
2006 Jun 20/21 & 3907      & 21:35 & 05:31 & 5    & 3461    \\
2009 Jun 22/23 & 5005      & 22:26 & 06:20 & 5    & 4950    \\
2009 Jun 23/24 & 5006      & 21:45 & 06:38 & 5    & 6067    \\
2012 Jun 12    & 6090      & 02:35 & 05:51 & 6    & 1352    \\
2012 Jun 13    & 6091      & 02:04 & 03:33 & 6    & \phantom{0}820 \\ [1ex]
2012 Jun 14/15 & 6093      & 23:47 & 01:15 & 6    & \phantom{0}770 \\
2012 Jun 15/16 & 6094      & 23:11 & 06:19 & 6    & 2691    \\  
2013 Jun 10/11 & 6454      & 23:28 & 05:08 & 5.5  & 2600    \\
2013 Jun 13/14 & 6457      & 22:10 & 08:03 & 5.5  & 5689    \\
\hline
\end{tabular}
\flushleft
\tablefoottext{a}{Light curve already used in Paper I}
\end{table}

Time-series imaging of the field around V893~Sco was performed
to generate light curves with a total duration of up to almost 9 hours.
Since the main purpose of the observations consisted in the determination
of eclipse timings, the characterization of rapid oscillations and the 
properties of the flickering activity (this
aspect is not pursued in the present paper) I aimed at a high time
resolution. In order to maximize the count rates
the observations were performed in white light, 
except for the night of 1999 June 11/12, when a CuSO$_4$ filter 
(transmitting basically the blue and ultraviolet light) as defined by 
Bessell (\cite{Bessell}) was used.

Until 2009 a thin back-illuminated EEV-CCD was employed. Its frame
transfer capabilities enabled observations with negligible dead time
between exposures. After this detector had been decommissioned
I used an Andor IXon camera. The small but not negligible 
read-out time of the detector was compensated by its superior efficiency.

The basic reductions of the data (de-biasing, flat-fielding) 
were performed using IRAF. Light curves prior to 2013 were constructed
from the direct images using the IRAF script LCURVE
(courtesy Marcos P. Diaz) which makes use of DAOPHOT/
APPHOT routines. An aperture size 2.2 times the FWHM
of the stellar images was chosen. On an absolute scale it varied according
to the seeing conditions (typically $2\arcsec - 2\farcs5$).
Four stars in the vicinity of V893~Sco were used as comparison stars. The
further analysis of the data was done using the MIRA software system 
(Bruch \cite{Bruch93}). 
MIRA was also used to construct the 2013 light curves after
a routine equivalent to LCURVE had been implemented.

\section{Photometric state and characteristics of the long-term light curve}
\label{Photometric state}

In order to assess the reliability of the relative magnitudes (the
difference between the instrumental magnitude of the variable star and the
principle comparison star $C_1$), the constancy of $C_1$ was checked, 
calculating the average difference between $C_1$ and other comparison stars
for all observing nights. Expect for two nights when this difference deviated
significantly from the mean for two different secondary comparison stars it
remained constant within 
$\pm 0^{\raisebox{.3ex}{\scriptsize m}}_{\raisebox{.6ex}{\hspace{.17em}.}}$03.
I can therefore be reasonable sure that $C_1$ is sufficiently stable to
a level that any residual variability does not affect the results presented
here.

The Second USNO CCD Astrograph Catalog (2UCAC) (Zacharias et al.\ 
\cite{Zacharias}) quotes a magnitude of 
12$^{\raisebox{.3ex}{\scriptsize m}}_{\raisebox{.6ex}{\hspace{.17em}.}}$85 for 
$C_1$ = 2UCAC~19933473 
in the bandpass 579--642~nm (in between V and R). While this number is not 
particularly accurate and may have an error of 
0$^{\raisebox{.3ex}{\scriptsize m}}_{\raisebox{.6ex}{\hspace{.17em}.}}$3, it can be used to
calculate the approximate visual magnitude of V893~Sco during the 
epochs of observations. 

\begin{figure}
   \centering
   \resizebox{\hsize}{!}{\includegraphics{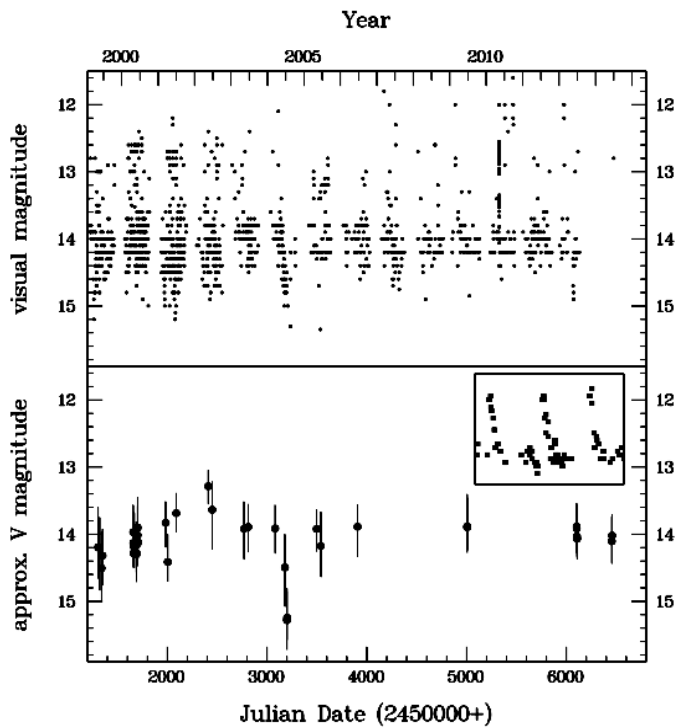}}
   \caption{{\it Top}: Visual AAVSO long-term light curve of V893~Sco. 
               {\it Bottom}: Average out-of-eclipse brightness of 
               V893~Sco 
               (transformed to approximate $V$-band magnitude) during each of
               the observing nights listed in 
               Tab.~\ref{Journal of observations}. The error bars represent
               the total range of magnitudes observed on the respective nights
               (disregarding eclipses). The insert shows on an extended scale
               a 65-day window, centred on JD 2451700, of the AAVSO light 
               curve with three distinct small outburst.}  
    \label{long-term-lc}
    \end{figure}

The AAVSO long-term light curve of V893~Sco
is shown in the upper panel of Fig.~1 (only positive observations
are plotted). The system remains most of the time at a visual magnitude of 
$\sim$$14^{\raisebox{.3ex}{\scriptsize m}}$, with frequent excursions to 
$\sim$$12^{\raisebox{.3ex}{\scriptsize m}}_{\raisebox{.6ex}{\hspace{.17em}.}}0$ -- 
$12^{\raisebox{.3ex}{\scriptsize m}}_{\raisebox{.6ex}{\hspace{.17em}.}}5$.
Interpreting these excursions as dwarf nova type outbursts, their 
amplitudes are only of the order of 
1$^{\raisebox{.3ex}{\scriptsize m}}_{\raisebox{.6ex}{\hspace{.17em}.}}$5 -- 
$2^{\raisebox{.3ex}{\scriptsize m}}$; significantly 
smaller than the outburst amplitudes of most dwarf novae. Occasionally, the
quiescent magnitude of V893~Sco drops to 
$\sim$$15^{\raisebox{.3ex}{\scriptsize m}}$. The lower panel of 
Fig.~\ref{long-term-lc} shows the mean approximate V magnitude of V893~Sco
at the epochs of the current observations. 
The error bars do not represent observational errors but indicate the total
(out-of-eclipse) range observed during a given night. Their size indicates
that an appreciable
part of the scatter in the AAVSO long-term visual light curve can be 
attributed to short term variations of the star. It is obvious from 
Fig.~\ref{long-term-lc}
that I observed V893~Sco almost exclusively during quiescence. Only on one
night (2002 May 15/16) was the system on the decline from an outburst 
(covered by only two data points in the AAVSO light curve). On two other 
nights (2004 July 13/14 and 14/15; not resolved in the figure) 
V893~Sco was in a low state, about 1$^{\raisebox{.3ex}{\scriptsize m}}$ --
1$^{\raisebox{.3ex}{\scriptsize m}}_{\raisebox{.6ex}{\hspace{.17em}.}}$5 fainter than 
normal quiescence. This coincides with a faint state in the AAVSO light curve.

The AAVSO light curve also shows that the long-term behaviour of 
V893~Sco is different from that of most dwarf novae below the period gap. There
are rather frequent low amplitude outbursts.
The insert in the lower frame of Fig.~\ref{long-term-lc} shows a 65-day time
window around JD~2451700 which contains 3 distinct outbursts. 
The recurrence time during this particular section of the light curve is 
about 20 days. But while
apart from the rather low amplitude the outburst behaviour is consistent to
what is seen on other dwarf novae, the fact that during 13 years of close
monitoring not a single superoutburst in this otherwise quite active system
has been seen is surprising. 

In order to assess the probability that superoutbursts have been missed 
by the AAVSO observers some simple simulations were performed. From the vast
collection of superoutburst light curves of SU~UMa stars published by Kato
et al.\ (\cite{Kato09}, \cite{Kato10}, \cite{Kato12}, \cite{Kato13}) it is
seen that the plateau phase lasts at least 10 days in the overwhelming 
majority of cases. I conservatively assume that V893~Sco should be 
brighter than 12$^{\raisebox{.3ex}{\scriptsize m}}_{\raisebox{.6ex}{\hspace{.17em}.}}$2 
during that phase. Regarding the distribution of observing epochs in the 
AAVSO light curve (now also considering negative observations if the limiting
magnitude is fainter than
12$^{\raisebox{.3ex}{\scriptsize m}}_{\raisebox{.6ex}{\hspace{.17em}.}}$2), and if V893~Sco
would have had a single superoutburst lasting 10 days starting some time
during the period covered by the AAVSO observers, the simulations ($10^6$ 
trials) showed that the likelihood for the superoutburst to have been missed is
$P_{\rm mis}(1) = 0.26$ (in 22\% of all cases due to seasonal gaps of 
$\approx$100 days in the light curve). This
probability drops rapidly with increasing duration of the outburst. If $N$ 
superoutburst would have occurred, to first order\footnote{If the 
superoutbursts would occur independently from each other, which they
do not!} the probability would be $P_{\rm mis}(N) = [P_{\rm mis}(1)]^N$. 
It would thus become very
small if the recurrence time of superoutbursts would be of the order of
1 year (or less) as typically observed in SU~UMa stars (unless the supercycle
happens to be close to one year or multiples thereof and all the 
superoutbursts manage to hide in the seasonal gaps in the AAVSO light curve).

In any case, practically all dwarf novae with
orbital periods below the 2-3~hour period gap are SU~UMa stars, 
but V893~Sco, so far, cannot yet unambiguously be classified as such.

\section{Eclipse profile}
\label{Eclipse profile}

As already noticed in \cite{PaperI}, the eclipses of V893~Sco 
are extremely variable. This refers equally to their depths, minimum
level and shapes as can be gauged from Figs.~2 and 3 of that study. This 
behaviour made the authors believe the hot spot (together with
part of the accretions disk) to be the eclipsed body, while the white
dwarf remains uneclipsed. However, in {\em Suzaku} observations
Mukai et al. (\cite{Mukai}) 
found a partial x-ray eclipse which they attributed to
a partial obscuration of the boundary layer between accretion disk and
white dwarf by the secondary star. This implies that the optical eclipse
should include at least a component due to the white dwarf being partially
covered by the red dwarf.

The additional optical data presented here greatly increase the number of 
observed eclipses, enabling -- in spite of the strong variability -- to
address the question of the eclipse shape in greater detail than was
possible before. In order to overcome the difficulty caused by the strong
variations of the eclipse depth the eclipses were first normalized. This is
possible because in most light
curves one can identify the start of eclipse ingress reasonably well even in 
the presence of strong flickering (see
Fig.~3 of \cite{PaperI}). In all cases this was done by visual inspection.
This introduces a certain degree of arbitrariness in particular when the
exact phase of ingress is difficult to separate from flickering activity.
However, while this increases the error margins, is does not affect the
conclusions. 

\begin{figure}
   \centering
   \resizebox{\hsize}{!}{\includegraphics{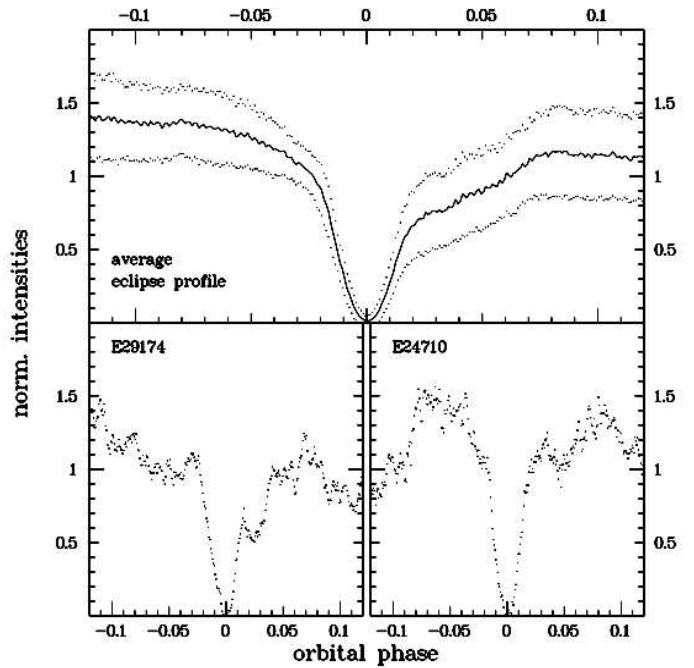}}
      \caption[]{{\em Top:} Average normalized eclipse profile 
                 of V893~Sco (solid line)
                 based on 110 individual eclipses. The dotted lines show the
                 standard deviation of individual profiles with respect to
                 the average. {\em Bottom:} Two individual eclipse profiles
                 with (left) and without (right) a clear delay of the final
                 egress.}
\label{ecl-profile}
\end{figure}

First, the light curves were phased according to the ephemerides derived in
Sect.~\ref{Ephemerides}. They were then transformed from 
magnitudes into arbitrary intensities and normalized such that the eclipse 
bottom, determined as the minimum of a fitted forth order polynomial to
the data points in the phase range from -0.01 to 0.01, has zero intensity,
and unit intensity corresponds to the difference between the brightness at the 
start of ingress and eclipse bottom. All eclipses are thus brought onto the 
same scale. The solid line in the upper frame of Fig.~\ref{ecl-profile} shows
the average profile of 110 eclipses\footnote{This number is smaller than
the total number of eclipses listed in Tab.~\ref{Eclipse epochs} because in
a few cases the light curve does not cover the start of eclipse ingress.}
used in this exercise with a resolution
of 0.001 in phase. The dotted lines delimit the uncertainty range as
determined from the standard deviation of all data points within a given
phase bin.

In many quiescent dwarf novae the ingress and egress phases of the eclipse 
profile consist of two components: Ingress starts with a steep brightness
decline ascribed to the occultation of the white dwarf. It is 
followed by a slightly slower decline caused by the eclipse of the bright spot.
When both, white dwarf and bright spot, are aligned along the line
of sight from the observer to the border of the secondary star at the start
of the eclipse, or if the transition is very smooth, 
the two components cannot be separated confidently.
On the other hand, upon egress the separation is often much clearer. The 
white dwarf emerges from the eclipse first. Due to the different geometrical
situation when looking along the opposite border of the red dwarf it takes
longer for the bright spot to emerge from the eclipse. Thus, egress often
occurs in two well separated steps. Classical examples of dwarf novae which
exhibit this kind of eclipse profile are Z~Cha (Wood et al., 
\cite{Wood86}), OY~Car (Wood et al., \cite{Wood89}) and 
IP~Peg (Bobinger et al., \cite{Bobinger}).

The average eclipse profile of V893~Sco shows some similarity to this classical
picture. The ingress cannot be separated into two distinct phases. 
It starts as phase $-0.019$ and may be interpreted as the 
simultaneous start of white dwarf and hot spot occultation. 
After a rounded minimum (indicating that the eclipse is not total) first a
steep rise can be seen which is symmetrical to the latter part of the ingress.
Apparently, this is due to the white dwarf emerging from partial eclipse.
After recovering to about 2/3 of the intensity at the start of ingress,  
the egress levels off at phase $0.018$ (i.e.\ within the errors this phase 
and the phase of the start of ingress are symmetrical to mid-eclipse) and 
continues much more gradually until the eclipse ends approximately at phase 
$0.077$. This second part of the egress may then be taken as the hot spot
egress. In contrast to what is seen in the examples quoted above there is no
distinct step at the end. This may either be explained 
by the step being washed out in the average of many eclipses or that the hot
spot has no well defined edge.

The error limits around the average eclipse profile, as shown in 
Fig.~\ref{ecl-profile}, are quite wide during the second part of the egress.
This reflects a strong variability of V893~Sco after the white dwarf 
emerges from occultation: In many cases the egress of the hot spot is not
evident in the individual light curves. To illustrate this, the lower frames
of Fig.~\ref{ecl-profile} show two eclipses, one with an obvious step during
egress (left) and one with no trace of a hot spot eclipse (right). This can
be explained by the strong ubiquitous flickering in V893~Sco which
apparently outshines the hot spot easily. In the case of the cataclysmic
variables Z~Cha, HT~Cas, V2051~Oph, 
IP~Peg and UX~UMa, Bruch (\cite{Bruch96}, \cite{Bruch00})
has shown that the flickering arises mainly in the inner accretion disk, 
albeit with contributions from the hot spot in most of these systems. The 
strong variability of V893~Sco during the second part of eclipse 
egress, i.e.\ when the hot spot is still hidden behind the secondary star, 
shows that in this case the hot spot is not the (main) flickering light source.

This scenario can also explain the strong variability of the eclipse depth
(see Fig.~2 of \cite{PaperI}). Since the eclipse of the white dwarf is only 
partial even at mid-eclipse about half of the inner accretion disk, where the 
flickering is expected to happen, remains visible. Since the short timescales 
of the Keplerian motion in the inner accretion disk ($\sim$$15^{\rm s}$ 
to complete an orbit at the surface of a typical $0.75 M_{\sun}$ white dwarf; 
$\sim$$165^{\rm s}$ sec at a distance of 5 white dwarf radii) inhibits 
flickering close to the white dwarf to maintain a significant 
asymmetry in azimuth
over the eclipse duration ($\sim$$4^{\rm m}$) the amount of eclipsed light 
and thus the eclipse amplitude should scale linearly with the out-of-eclipse
brightness, and the factor of proportionality should be equal to the fraction 
of light occulted at mid eclipse. 

   \begin{figure}
   \centering
   \resizebox{\hsize}{!}{\includegraphics{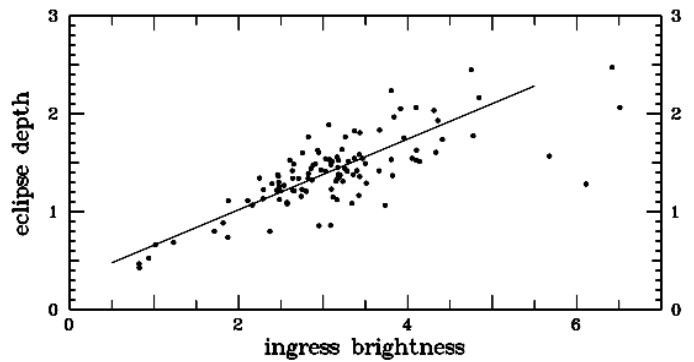}}
      \caption[]{Eclipse depth as a function of the brightness of 
                 V893~Sco at the start of eclipse ingress, both in 
                 arbitrary intensity units. The solid line represents a 
                 linear fit to 
                 the data, disregarding the four deviating points at the right
                 which refer to an outburst (see text).}
\label{ecl-depth}
\end{figure}

Fig.~\ref{ecl-depth} shows the dependence of the eclipse depth on the
brightness at the start of eclipse ingress. Again, magnitudes have
been transformed into arbitrary intensities which, however, in the present
context must not be normalized.
There is a clear correlation, in spite of a significant scatter
which may be attributed partly to difficulties to exactly define the
onset of the eclipses in the individual light curves and partly to 
uncertainties caused by a residual dependence of flickering on azimuth
in the inner disk. The four deviating points at high ingress brightness
all refer to eclipses observed on 2002, May 15/16, when V893~Sco 
was on the decline from an outburst (see Sec.~\ref{Photometric state}). 
The disk structure should therefore be different during these eclipses and  
I disregard these points.
A linear fit to the remaining data, shown as a solid line in the figure, 
has a slope of $0.36 \pm 0.06$, suggesting that on the average about
a third of the light of V893~Sco is occulted during 
mid-eclipse. 

\section{Ephemerides}
\label{Ephemerides}

Eclipse ephemerides for V893~Sco have first been derived in 
\cite{PaperI} using a total time base of only 49 days. They
have accumulated a significant error over time. Therefore 
this exercise is repeated here with higher precision, 
taking advantage of the much longer time base of 14 years.

The same method to define the eclipse timings was used here as employed 
in \cite{PaperI} (the reader is referred to that paper for details) 
The eclipse epochs (including also the eclipses analysed
in \cite{PaperI}\footnote{Except those of 1999 May 4 and 5 which were observed
with much lower time resolution so that the mentioned methods to determine the
eclipse epochs could not be applied; see \cite{PaperI}.}) , 
expressed as Barycentric 
Julian Date in Barycentric Dynamical Time (BJD-TDB)\footnote{Transformation 
from UTC as given by the Telescope Control System
during observations to Barycentric Dynamical Time was performed using the
tool available at http://astroutils.astronomy.ohio-state.edu/time/ (see Eastman
et al., \cite{Eastman}).} are
given in Table~\ref{Eclipse epochs}. I disregard here the data of 1999 June 
21/22, because the eclipse times deviate strongly from the
ephemerides derived below, suggesting that a timing error occurred during this
night. Similarly, an error of the time stamps of a part of the 2013 June 10/11,
observation inhibited the use of the third eclipse observed during that 
night.

\begin{table*}
\caption{Eclipse Epochs.}
\label{Eclipse epochs}
\centering

\begin{tabular}{l@{\hspace{1ex}}c@{\hspace{1ex}}c@{\hspace{1ex}}|
l@{\hspace{1ex}}c@{\hspace{1ex}}c@{\hspace{1ex}}|
l@{\hspace{1ex}}c@{\hspace{1ex}}c@{\hspace{1ex}}}
\hline\hline
  & Eclipse  & BJD-TBD   &   & Eclipse  & BJD-TBD  &  & Eclipse  & BJD-TBD \\
Date & Number   & (2450000+) & 
Date & Number   & (2450000+) &
Date & Number   & (2450000+) \\
\hline 
1999 May 25 & \phantom{0000}0 & 1323.53330 & 
2000 Jun  7 & \phantom{0}4992 & 1702.73250 & 
2004 Jun 22 &           24433 & 3179.49968 \\
   
            & \phantom{0000}1 & 1323.60957 & 
            & \phantom{0}5001 & 1703.41681 & 
2004 Jun 23 &           24434 & 3179.57572 \\
   
            & \phantom{0000}3 & 1323.76137 & 
            & \phantom{0}5002 & 1703.49254 & 
            &           24435 & 3179.65176 \\
   
1999 Jun 11 & \phantom{00}236 & 1341.46053 & 
2000 Jun  8 & \phantom{0}5003 & 1703.56856 & 
            &           24436 & 3179.72767 \\
   
1999 Jun 12 & \phantom{00}237 & 1341.53637 & 
            & \phantom{0}5004 & 1703.64424 & 
2004 Jul 13 &           24709 & 3200.46521 \\
   
            & \phantom{00}238 & 1341.61232 & 
            & \phantom{0}5005 & 1703.72061 & 
2004 Jul 14 &           24710 & 3200.54117 \\
   
            & \phantom{00}240 & 1341.76433 & 
            & \phantom{0}5006 & 1703.79636 & 
            &           24711 & 3200.61721 \\
   
2000 Apr 25 & \phantom{0}4424 & 1659.58695 & 
            & \phantom{0}5014 & 1704.40419 & 
            &           24722 & 3201.45257 \\
   
2000 Apr 26 & \phantom{0}4437 & 1660.57400 & 
            & \phantom{0}5015 & 1704.48013 & 
2004 Jul 15 &           24723 & 3201.52871 \\
   
            & \phantom{0}4438 & 1660.65015 & 
2000 Jun  9 & \phantom{0}5016 & 1704.55593 & 
2005 May  7 &           28621 & 3497.62663 \\
   
2000 Apr 27 & \phantom{0}4451 & 1661.63786 & 
            & \phantom{0}5017 & 1704.63210 & 
            &           28622 & 3497.70267 \\
   
2000 Apr 28 & \phantom{0}4464 & 1662.62522 & 
            & \phantom{0}5018 & 1704.70801 & 
            &           28623 & 3497.77880 \\
   
2000 May  6 & \phantom{0}4568 & 1670.52540 & 
            & \phantom{0}5019 & 1704.78413 & 
2005 Jun 17 &           29172 & 3539.48133 \\
   
            & \phantom{0}4569 & 1670.60127 & 
2000 Jun 10 & \phantom{0}5029 & 1705.54367 & 
2005 Jun 18 &           29174 & 3539.63342 \\
   
2000 May 23 & \phantom{0}4792 & 1687.54082 & 
            & \phantom{0}5030 & 1705.61962 & 
            &           29175 & 3539.70938 \\
   
            & \phantom{0}4794 & 1687.69242 & 
            & \phantom{0}5031 & 1705.69550 & 
2006 Jun 20 &           34016 & 3907.43862 \\
   
            & \phantom{0}4795 & 1687.76871 & 
            & \phantom{0}5032 & 1705.77137 & 
2006 Jun 21 &           34017 & 3907.51460 \\
   
            & \phantom{0}4796 & 1687.84446 & 
2001 Mar 16 & \phantom{0}8705 & 1984.77784 & 
2009 Jun 22 &           48471 & 5005.46165 \\
   
2000 May 24 & \phantom{0}4805 & 1688.52834 & 
2001 Apr  6 & \phantom{0}8979 & 2005.59115 & 
2009 Jun 23 &           48472 & 5005.53778 \\
   
            & \phantom{0}4806 & 1688.60412 & 
2001 Jun 30 &           10097 & 2090.51631 & 
            &           48473 & 5005.61377 \\
   
            & \phantom{0}4807 & 1688.68012 & 
            &           10098 & 2090.59219 & 
            &           48474 & 5005.68957 \\
   
            & \phantom{0}4809 & 1688.83201 & 
            &           10099 & 2090.66788 & 
            &           48475 & 5005.76538 \\
   
2000 May 25 & \phantom{0}4818 & 1689.51563 & 
2002 May 16 &           14310 & 2410.54171 & 
            &           48484 & 5006.44943 \\
   
            & \phantom{0}4819 & 1689.59161 & 
            &           14311 & 2410.61774 & 
2009 Jun 24 &           48485 & 5006.52508 \\
   
            & \phantom{0}4820 & 1689.66779 & 
            &           14312 & 2410.69363 & 
            &           48486 & 5006.60131 \\
   
            & \phantom{0}4821 & 1689.74354 & 
            &           14313 & 2410.76953 & 
            &           48487 & 5006.67710 \\
   
            & \phantom{0}4822 & 1689.81953 & 
2002 Jun 24 &           14836 & 2450.49766 & 
            &           48488 & 5006.75309 \\
   
2000 May 26 & \phantom{0}4831 & 1690.50306 & 
2002 Jun 25 &           14837 & 2450.57328 & 
2012 Jun 12 &           62757 & 6090.64690 \\
   
            & \phantom{0}4832 & 1690.57923 & 
            &           14838 & 2450.64943 & 
2012 Jun 13 &           62770 & 6091.63449 \\
   
            & \phantom{0}4834 & 1690.73105 & 
            &           14839 & 2450.72537 & 
2012 Jun 15 &           62795 & 6093.53328 \\
   
2000 May 27 & \phantom{0}4845 & 1691.56675 & 
2003 May 10 &           19036 & 2769.53553 & 
2012 Jun 16 &           62808 & 6094.52083 \\
   
            & \phantom{0}4846 & 1691.64260 & 
            &           19037 & 2769.61168 & 
            &           62811 & 6094.74881 \\
   
            & \phantom{0}4847 & 1691.71863 & 
2003 Jun 20 &           19588 & 2811.46646 & 
2013 Jun 11 &           67547 & 6454.50234 \\
   
            & \phantom{0}4848 & 1691.79459 & 
2003 Jun 21 &           19589 & 2811.54245 & 
            &           67548 & 6454.57837 \\
   
2000 Jun  6 & \phantom{0}4988 & 1702.42887 & 
            &           19590 & 2811.61860 & 
2013 Jun 13 &           67586 & 6457.46487 \\
   
2000 Jun  7 & \phantom{0}4989 & 1702.50459 & 
            &           19591 & 2811.69426 & 
2013 Jun 14 &           67587 & 6457.54111 \\
   
            & \phantom{0}4990 & 1702.58067 & 
2004 Mar 17 &           23145 & 3081.66173 & 
            &           67589 & 6457.69266 \\
   
            & \phantom{0}4991 & 1702.65665 & 
            &           23147 & 3081.81294 & 
            &           67590 & 6457.76880 \\
   
\hline
\end{tabular}
\end{table*}

It is not straight forward to assign error bars to the individual eclipse 
epochs. Requiring that the reduced $\chi^2_{\rm r}$ of the residuals between the
observed and calculated minimum times, using the final ephemerides (see below),
is equal to 1 (i.e.\ that the minimum times are adequately described by
eq.~\ref{eq. 1}) results in an error of 13.3~sec which is adopted a
posteriori as the standard deviation of the data points.

\begin{figure}
   \centering
   \resizebox{\hsize}{!}{\includegraphics{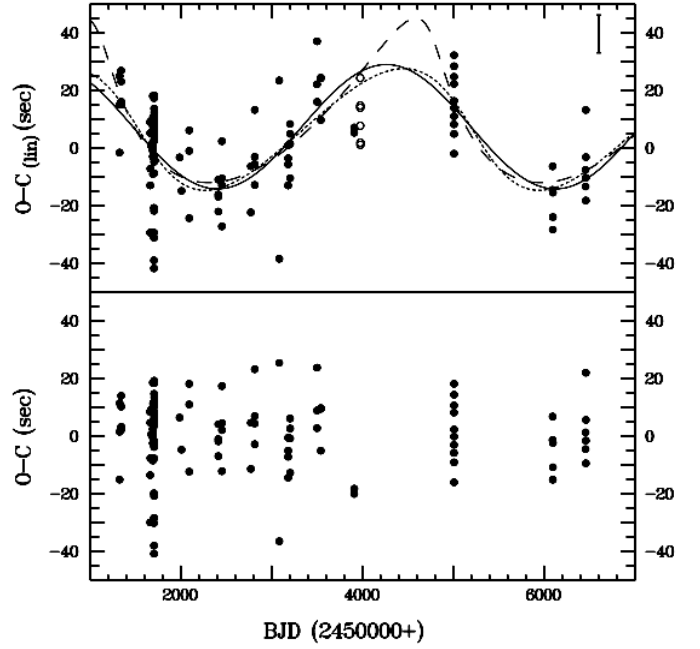}}
      \caption[]{{\em Top:} 
                 Observed minus calculated eclipse epochs for 
                 V893~Sco, using
                 the best fit linear ephemerides. The filled and open symbols 
                 represent eclipses of this paper and of Mukai et al. (2009),
                 respectively. The solid line is the best fit sine curve to
                 the filled symbols. The dotted and dashed curves represent
                 solutions for an eccentric orbit of an assumed third body
                 with eccentricity $e = 0.3$ and $e = 0.6$, respectively
                 (see Sect.~\ref{The third body: Basic parameters}).
                 An empirical estimate for the error
                 bars of the individual points is shown in the upper right 
                 corner.
                 {\em Bottom:}
                 Observed minus calculated eclipse epochs using the final
                 ephemerides described by Eq.~\ref{eq. 1}. 
}
\label{o-c-diagram}
\end{figure}

A linear least-squares fit to the eclipse timings listed in 
Table~\ref{Eclipse epochs} yields the $O-C$ diagram shown in 
the upper frame of Figure~\ref{o-c-diagram}. It clearly shows that 
linear ephemerides cannot adequately describe the minimum times of 
V893~Sco. There is an obvious modulation of the $O-C$ values 
which can very well be fitted with a simple sine-curve. 
Therefore, a suitable description of the ephemerides takes the form
\begin{equation}
\label{eq. 1}
BJD(min) = T_o + P_o \times E + b \times E^2 + 
           A\, \sin \left[2\pi \left(\frac{E}{P_1}-\phi \right)\right] \, ,
\end{equation}
where $BJD$ is the Barycentric Julian Date expressed in 
Barycentric Dynamical Time and $E$ is the eclipse number. 
Here, I also include a term $b \times E^2$,
allowing for a secular variation of the orbital period. $b$ is related to
the time derivative of $P_o$ through ${\dot P_o} \equiv {\rm d}P_o/{\rm d}t
= 2b/P_o$.

Numerical values for the time $T_o$ of 
eclipse $E=0$, the mean orbital period $P_o$ and its derivative 
${\rm d}P_o/{\rm d}t$, the period $P_1$ of the 
modulation of the orbit, the amplitude $A$ of the modulation and the phase 
$\phi$ are summarized in Table~\ref{Ephemeris-tab} (Model A) 
where the quantities in 
brackets represent the errors in the last digits. The $O-C$ diagram
corresponding the these final ephemerides is shown in the lower frame of
Figure~\ref{o-c-diagram}. It is consistent with 0 and does not show any 
significant systematic trend over the 14 year time base. Some points
close to BJD 2451700 are systematically below most of the other data. They
all refer to an observing run in 2000 June. While I cannot exclude an error
in the time stamps of these observations, there is also no clue for this to
be the case. Therefore, there is no objective reason to exclude these data
from the analysis.

\begin{table}
\caption{Ephemerides for V893~Sco.}
\label{Ephemeris-tab}
\centering

\begin{tabular}{ll@{\hspace{2ex}}r@{\hspace{2ex}}r}
\hline\hline
      &        & Model A             & Model B             \\
\hline
$T_o$ & (BJD)  & 2451323.533413 (15) & 2451323.533433 (15) \\ 
$P_o$ & (days) & 0.07596146514 (52)  & 0.07596146589 (53)  \\
\ldots & (hours) & 1.823075163 (13)  & 1.823075181 (13)    \\
${\rm d}P_o/{\rm d}t$ & & $6 (3) \, 10^{-13}$ & --          \\
$P_1$ & (cycles) & 48992 (1102)      & 48989 (1259)        \\
\ldots & (days) & 3721 (84)          & 3721 (95)           \\       
$A$ &(days)   & 0.000258 (27)        & 0.000258 (26)       \\ 
\ldots & (sec) & 22.3 (2.3)          & 22.3 (2.3)          \\
$\phi$ &    & 0.537 (11)             & 0.537 (50)          \\
\hline
\end{tabular}
\end{table}

While the period is well determined this is not so for the
period derivative. Nominally, V893~Sco exhibits a slight secular 
increase in the orbital period. While there may be
mechanisms which lead to a period increase over the scale of the
observational time base this is contrary to the expected secular period
evolution of cataclysmic variables unless V893~Sco is a period bouncer.
This, however, is rather unlikely since it does not exhibit any behaviour
considered characteristic for CVs which have evolved beyond the
period minimum (Zharikov et al., \cite{Zharikov}) such as broad Balmer 
line absorption troughs (see spectra shown by Mason et al. \cite{Mason}) 
or persistent orbital double humps in the light curve. Moreover, 
V893~Sco exhibits rather frequent (normal)
dwarf nova outbursts, in contrast to the rare (WZ~Sge type) superoutbursts
expected for period bouncers. Therefore, considering the small numerical value
of ${\rm d}P_o/{\rm d}t$ and the large relative error, $\dot{P_o}$ can at most 
be considered marginally significant and I will neglect it in the following.

Meeting a concern raised by the referee that the parameters $A$ and $b$ 
in eq.~\ref{eq. 1} are likely to be highly coupled I also calculated a 
solution with $b$ fixed to zero (Model B in 
Tab.~\ref{Ephemeris-tab}). All other parameters change only 
insignificantly (much less than their errors) with respect to the values 
for Model A which were used for the subsequent calculations. 
But Model B would lead to identical results within the
numerical accuracy of all quantities quoted below.

When recalculating the ephemerides I did not consider the eclipse timings
measured by Mukai et al. (\cite{Mukai}) 
because they determined eclipse ingress and
egress times in a way such that the eclipse centre (calculated as the mean 
between ingress and egress) might be systematically offset from the eclipse
centre as determined here. However, as a consistency check the results of
their timings, after transforming the heliocentric Julian Dates quoted by
them into barycentric Julian Dates, are included in the $O-C$ diagram of 
Fig.~\ref{o-c-diagram} (upper frame)
as open symbols. It is seen that they fit in quite satisfactorily.  

\section{The cyclic period variations}
\label{The cyclic period variations}

Cyclic period variations derived from eclipse timings are not unusual in 
cataclysmic variables. On the contrary, they appear rather to be the rule. 
Borges et al. (\cite{Borges}) compiled
a list of 14 systems exhibiting this phenomenon which contains most of the
brighter eclipsing CVs, well observed over a long time base. An updated
version was published by Pilar\v{c}\'{i}k et al.\ (\cite{Pilarcik}), 
who also included some
post-common envelope binaries (PCEBs) with similar period variations. 
HU~Aqr (Qian et al.\ \cite{Qian11}, 
                 Go\'zdziewskiet al. \cite{Gozdziewski}) and
RR~Cae (Qian et al.\ \cite{Qian12})
may be added to this list.

Discarding apsidal motions as the origin of the period variations on the
grounds that the orbital eccentricity of these strongly interacting binaries 
must be essentially zero, it was for a long time believed that they could be 
explained by magnetic cycles in the secondary stars of these systems via 
the so-called Applegate mechanism (Applegate \cite{Applegate92}).
However, unless a significantly more energy efficient mechanism can be 
identified (see Lanza et al.\ \cite{Lanza}) this explanation
has been shown not to work in many individual cases (e.g.\
AC~Cnc: Qian et al.\ \cite{Qian07}; 
Z~Cha: Dai et al.\ \cite{Dai09b};
TV~Col: Dai et al.\ \cite{Dai10b};  
UZ~For: Potter et al.\ \cite{Potter};
DQ~Her: Dai \& Qian \cite{Dai09a};
NN~Ser: Brinkworth et al.\ \cite{Brinkworth};
QS~Vir: Parsons et al.\ \cite{Parsons}) 
mainly because the energy required to drive
the observed period change is more than the total energy output of the star.

Therefore, researchers started to focus on light travel time effects to
explain the period variations. In this scenario the binary rotates around
the centre of gravity with a third object. Depending on its distance from
the centre of mass eclipses will be observed slightly early or late and 
thus the orbital period will appear to be variable with the period of this 
extra motion of the binary. Under this hypotheses the observed cyclic period
changes can in many cases be explained by the presence of a giant planet
or a brown dwarf orbiting around the binary 
(T~Aur: Dai \& Qian \cite{Dai10a};
RR~Cae: Qian et al.\ \cite{Qian12};
DQ~Her: Dai \& Qian \cite{Dai09a};
DP~Leo: Qian et al.\ \cite{Qian10a}, 
                 Beuermann et al.\ \cite{Beuermann11};
QS~Vir: Qian et al.\ \cite{Qian10b};
HS0705+6700: Beuermann et al.\ \cite{Beuermann12a}).

Adopting this basic hypotheses the interpretation of the observations is 
relatively straightforward in the quoted cases. However there are other 
examples where the period variations are more complex and two planets (and, 
additionally, often a secular period decrease) 
are required to explain them within the light travel time
scenario. The question of long-term stability of such planetary systems then
arises, and it may not be easy (or even impossible) to find a configuration
which is compatible with observations and at the same time ensures the
survival of the planetary system on secular timescales. Examples are 
HU~Aqr (Qian et al.\ \cite{Qian11}, Horner et al.\ \cite{Horner11}, 
                 Go\'zdziewski et al.\ \cite{Gozdziewski}
                 Wittenmyer et al.\ \cite{Wittenmyer}),
NN~Ser (Qian et al.\ \cite{Qian09}, 
                 Brinkworth et al.\ \cite{Brinkworth}, 
                 Beuermann et al.\ \cite{Beuermann10}, 
                 Hessman et al.\ \cite{Hessman}, 
                 Marsh et al.\ \cite{Marsh}) and
HW~Vir (Lee et al.\ \cite{Lee}, 
                 Beuermann et al.\ \cite{Beuermann12b}, 
                 Horner et al.\ \cite{Horner12}).

Another important requirement for the light travel time interpretation to be 
viable arises from evolutionary considerations. The orbit(s) must
be such that the planet(s) not only can survive the common-envelope phase, but
they must also have been 
stable in the previous phase when the stellar components had
a much larger separation. A related question is the fraction of PCEBs which 
show evidence for the presence of planets as compared to the fraction of 
their progenitors with planetary systems. Zorotovic and Schreiber 
(\cite{Zorotovic})
find that 90\% of detached PCEBs with accurate eclipse time measurements
show period variations that might indicate the presence of a third body.
However, their main sequence progenitors are quite unlikely to be hosts of
giant planets ($\le 10\%$). These authors therefore conclude it to be more
likely that the period variations in PCEBs are caused by second generation
planets, formed from remnants of the common envelope which were not expelled
from the binary system, or that they are due to some other cause altogether
which so far has not yet been identified.

Finally, the question of the uniqueness of a given solution must be addressed.
More than once a published solution was shown to be untenable or required
a significant revision when more 
eclipse timings were added to the available data
(e.g.\ DP~Leo: Qian et al.\ \cite{Qian10a} vs.\ 
                       Beuermann et al.\ \cite{Beuermann11};
NN~Ser: Qian et al.\ \cite{Qian09} vs.\ 
                 Parsons et al.\ \cite{Parsons} and 
                 Beuermann et al.\ \cite{Beuermann10};
HW~Vir: Lee et al.\ \cite{Lee} vs.\ 
                 Beuermann et al.\ \cite{Beuermann12b}). 
Independent analysis of the same data set, using different criteria, 
may also lead to discrepant results (e.g.\ 
HU~Aqr: Qian et al.\ \cite{Qian11} vs.\ 
                 Go\'zdziewski et al.\ \cite{Gozdziewski};
HW~Vir: Lee et al.\ \cite{Lee} vs.\ Horner et al.\ \cite{Horner12}).
Understandably this problem is more severe for systems with 
complex period variations which cannot be explained by the effect of a 
single planet.
 
In contrast to many of the cases quoted above the period variations of 
V893~Sco are comparatively simple and can well be approximated by a 
single sine curve. In spite of the mentioned caveats it is therefore 
close at hand to interpret them as a
light travel time effect caused by a third body in the system; and I will do
so in the following. However, first I will verify if the Applegate mechanism
can safely be rejected.

\subsection{The Applegate mechanism}
\label{The Applegate mechanism}

As mentioned above it has been shown that the Applegate
mechanism cannot explain the period variations observed in several 
CVs and pre-CVs because the secondary star does not generate enough
energy to drive it. Here, I will briefly verify if this is also the
case for V893~Sco.

In order to explain the full amplitude $2 A$ of the observed $O-C$ variations
over a time span of $P_1/2$ an average period derivative of $\delta P/P_0 =
1.44 \times 10^{-7}$ is required. This implies in a total period change of
$\Delta P = 3.5 \times 10^{-3}\, {\rm s}$. 
 
For an approximate calculation of the total energy generated by the 
secondary star I assume that its radius $R_2$ and effective temperature 
$T_2$ are equal to the values derived from the semi-empirical CV donor 
star sequence of Knigge et al. (\cite{Knigge11}) 
at the orbital period of V893~Sco. This gives 
$R_2 = 0.174\, R_{\sun}$ and $T_2 = 3087$ K. The luminosity is then
$L_2 = 4 \pi R_2^2 \sigma T_2^4 = 2.47\times 10^{-3}\, L_{\sun}$ where $\sigma$ 
is the Stefan-Boltzmann
constant. Over the time required to change the orbital period by $\Delta P$ 
(i.e.\ half the period of the cyclic variation of the eclipse timings)
this adds up to $1.5\times 10^{39}\, {\rm erg}$. 

On the other hand, using eqs.\ 27 and 28 of Applegate (\cite{Applegate92}) 
the energy
required to redistribute the angular momentum between the core and the shell
of the star can be calculated. Being interested in the minimum of the
necessary energy I neglect the angular velocity of differential rotation in
Applegate's eq.\ 28: $\Omega_{\rm dr} = \Omega_{\rm shell} - \Omega_{\rm core} = 0$.
Assuming component masses of $M_2 = 0.134\, M_{\sun}$ and $M_1 = 0.75\, 
M_{\sun}$ (see Sect.~\ref{The third body: Basic parameters}) for 
V893~Sco, the component separation follows from the orbital 
period and Kepler's third law. Furthermore, a simple polytrope model with 
index $n=1.5$, appropriate for fully convective stars, is used 
to calculate the interior mass distribution and the
moments of inertia of the core and the shell as a function of the assumed
shell mass. The minimum energy is then found to be $1.8\times 10^{39} {\rm erg}$
at a shell mass of $0.035\, M_{\sun}$. Although not by a large factor, this is 
still more than the total energy generated by the star. 

Since it cannot be
assumed that the entire energy production is used to power the
angular momentum distribution between layers of the star, Applegate's 
mechanism can thus be excluded as an explanation for
the cyclic period change observed in V893~Sco, unless an alternative 
energy source can be identified. 

Following a suggestion of Raymundo
Baptista (private communication) I make the {\em ad hoc} assumption that
orbital energy of the secondary can be tapped to drive the Applegate 
mechanism (without trying to specify a physical mechanism for this to happen).
Lowering the secondary star in the gravitational potential of the primary
will release
potential energy. Assuming Keplerian orbits, a part of this will be used up
for the faster motion of the secondary on the lower orbit, but the rest will 
be available for other purposes. I therefore calculated the change 
$\Delta P$
in the orbital period required for this rest to be equal to the minimum
energy necessary to drive the Applegate mechanism. Adopting the parameters
of V893~Sco, $\Delta P = -8.2\, 10^{-5}$~sec. Assuming that 
this energy must become available over the time it takes to complete half of 
a cycle of the period variations this translates into a time derivative of 
${\rm d}P_o/{\rm d}t = -5.1\, 10^{-13}$. This is of the order of the
error of the formal time derivative quoted in Tab.~\ref{Ephemeris-tab}. 
The cumulative
effect on the $O-C$ diagram over the whole time base of the observations
would be $\approx$1~sec, undetectable in
view of the scatter of the data points in Fig.~\ref{o-c-diagram}.

The picture drawn here is, of course, greatly simplified .It assumes that 
the entire available orbital energy is used to power the Applegate mechanism. 
If only a part of it goes into this channel $\Delta P$, as derived 
above, is a lower limit. On the other hand, apart from
orbital energy other energy sources may be accessible.
Finally, the necessary
energy for the Applegate mechanism adopted here is the minimum value:
If the shell mass is different from $0.035\, M_{\sun}$ or if $\Omega_{\rm dr} > 0$
more energy is required.
There are thus at least three unknown parameters which complicate the picture,
not to speak of the total ignorance of a physical mechanism to explain
how orbital energy can be tapped. 

Therefore, in conclusion it can be stated that under favourable circumstances
a secular period decrease smaller than can be detected with the available
data is, in principle, able to provide the energy necessary to drive the
Applegate mechanism. But, for the time being, this possibility is only 
speculative.

\subsection{Basic parameters of the third body}
\label{The third body: Basic parameters}

Therefore, I will now interpret the cyclic variations in the $O-C$ diagram 
as being caused by the motion of a third body.
Subsequently the indices ${\cal A}$ and ${\cal B}$ will be used for the white 
dwarf and the mass transferring red dwarf in V893~Sco, and 
${\cal P}$\footnote{anticipating that it may be a planet (instead of, e.g.\
a brown dwarf).} for the 
third body. The index ``${\rm inn}$'' refers to the inner binary, composed of 
components ${\cal A}$ and ${\cal B}$, while ``${\rm out}$'' refers to the outer 
orbit of ${\cal P}$ around ${\cal AB}$.

The shape of the $O-C$ curve does not suggest significant deviations from a
pure sinusoidal signal, and the scatter of the data points does not warrant a
fit of a more complicated function such as that needed to describe an eccentric
orbit. Therefore, the motion of the third body around the inner V893~Sco 
binary is assumed to be circular. Let $c$ be the speed of light and 
$i_{\rm out}$ the inclination of the outer orbit to the line of sight.
Then the amplitude $A$ of the eclipse timings modulation translates into a 
projected distance of component ${\cal AB}$ to the centre of gravity of 
$a_{\cal AB} \sin i_{\rm out} = A \times c
= (6.7 \pm 0.7) \times 10^9 \, {\rm m}$. With the total mass of the
inner binary system, $M_{\cal AB}$, the mass of the third body, $M_{\cal P}$,
the modulation period $P_1$, and the gravitational constant $G$, 
the mass function can be written as 
  
\begin{eqnarray}
f(m) \equiv \frac{\left( M_{\cal P} \sin i_{\rm out}\right)^3}
                 {\left( M_{\cal AB} + M_{\cal P} \right)^2} 
        &  =  & \frac{4 \pi^2}{G}
            \frac{\left( a_{\cal AB} \sin i_{\rm out}\right)^3}{P_1^2}
            \nonumber \\
      &  =  & \left( 1.7 \pm 0.5 \right) \, 10^{24}\, {\rm kg} \, . 
\label{eq. 2}
\end{eqnarray}

\noindent In order to get a handle on the mass of the third body from this 
equation, information
about the inclination of the outer orbit and the mass of the inner binary is
required. For both quantities no direct measurements exist.

Based on the empirical relation between the mass of the donor stars in
cataclysmic variables and the binary
period (Warner, \cite{Warner95}), Matsumoto et al.\ (2000) 
estimate a mass of $M_{\cal B} = 0.15 \, M_{\sun}$ for V893~Sco. 
Together with the 
constrain on the orbital inclination $i_{\rm inn} \ga 70\degr$ imposed by the
presence of eclipses in the light curve and the amplitude of the radial
velocity curve of the $H\alpha$ emission line (assuming it to reflect the
motion of the white dwarf) they estimate 
$M_{\cal A} \approx 0.5 - 0.6 \, M_{\sun}$.
Mason et al. (\cite{Mason}) used the secondary star mass -- period and 
mass -- radius relations of Howell \& Skidmore (\cite{Howell}) 
to derive $M_{\cal B} = 0.175\, M_{\sun}$
and $R_{\cal B} = 0.196\, R_{\sun}$. They then iteratively used relations
involving $M_{\cal A}$, $M_{\cal B}$, $R_{\cal A}$, $R_{\cal B}$, $i_{\rm inn}$,
the accretion disk radius, and the bright spot radius and azimuth to
get $M_{\cal A} = 0.89\, M_{\sun}$. The secondary star mass of 
$M_{\cal B} = 0.15 \, M_{\sun}$ adopted by Mukai et al. (\cite{Mukai}) is based 
on Patterson's (\cite{Patterson}) empirical mass-period relationship.
They then assumed a mass ratio for the inner binary of $q_{\cal A,B} =
0.25$ to obtain a white dwarf mass of $M_{\cal A} = 0.58 \, M_{\sun}$\footnote{The
slight inconsistency between these numbers can be explained by rounding
errors.}. A slightly lower value of $M_{\cal B} = 0.134\, M_{\sun}$ for the 
secondary star mass follows from the semi-empirical CV donor star sequence
of Knigge et al. (\cite{Knigge11}). Knigge (\cite{Knigge06}) 
considers $M_{\cal A} = 0.75\, M_{\sun}$
as the mass of a white dwarf in a typical cataclysmic variable.

While the differences in the various $M_{\cal B}$-values are small and not
critical in the present context, all quoted values of the primary star mass
are based on uncertain assumptions. Here, I will adopt the mass
$M_{\cal A} = 0.75\, M_{\sun}$ of Knigge ({\cite{Knigge06}), 
assigning a generous error bar
of $0.25\, M_{\sun}$ to $M_{\cal A}$ which encompasses the range of masses quoted
in the literature. Since in the following calculation only the sum
$M_{\cal AB} = M_{\cal A} + M_{\cal B}$ enters this error is assumed also to
account for the much smaller error of the secondary star mass. Thus, 
$M_{\cal AB} = (0.88 \pm 0.25) \, M_{\sun}$. 

A relation between the inclination of the outer orbit and the mass of the
third body can then be constructed which is shown in Fig \ref{i-m3-relation}
where the mass is expressed in units of Jupiter masses. The dashed lines
indicate the $1 \sigma$ error limits.

   \begin{figure}
   \centering
   \resizebox{\hsize}{!}{\includegraphics{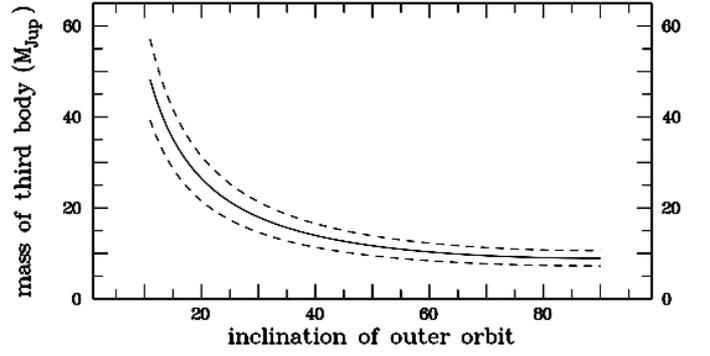}}
      \caption[]{Mass of the third body as a function of the inclination of
                 the outer orbit to the line of sight. The dashed curves 
                 indicate the $1 \sigma$ error limits.}
\label{i-m3-relation}
\end{figure}

Knowing the orbital period of the third body and the component masses
Kepler's third law yields immediately its distance to the central binary:
$a_{\rm out} = {4.5 \pm 0.4} \, {\rm AU}$. Due to the dependence of the mass of 
the third body on $i_{\rm out}$ there is a very slight increase with the 
inclination of the orbit. This is, however, wholly negligible within the 
errors which are strongly dominated by the uncertainty of the white dwarf mass.

If the original binary star which developed into
a CV and the third body were formed via fragmentation of the same rotating
interstellar cloud, or if the third body coalesced from a circum-binary disk
(and if the orbital inclination did not suffer from a significant perturbation
during the subsequent evolution) the angular momentum of the inner and outer 
orbits should be approximately aligned. Thus, $i_{\rm out} \approx i_{\rm inn}$. 
The presence of eclipses prove V893~Sco to have a high inclination. 
Mason et al.\ (\cite{Mason}) estimate it to be $72\fdg5$, while Mukai et al.\ 
(\cite{Mukai}) derive a slightly higher value of $74\fdg2$. 
Fig.~\ref{i-m3-relation} shows that in this region the relation between the 
mass of the third body and the inclination of the outer orbit is almost flat 
and $M_{\cal P} \approx 9.5\, M_{\rm Jup}$. Therefore, if the inner and
outer orbits are in fact aligned, the third body has a mass which would
classify it as a giant planet. I will therefore subsequently not 
distinguish between the terms ``third body'' and ``planet''. 

In the previous discussion
the assumption of a circular orbit of the third body was justified 
with the scatter of the data points in the $O-C$ diagram. While the data
thus do not suggest a finite orbital excentricity $e$, they also do not
exclude values of $e > 0$. Respective solutions were investigated,
using the prescriptions of Irwin (\cite{Irwin}). A formal fit of an
eccentric orbit to the $O-C$ values shown in the upper frame of 
Fig.~\ref{o-c-diagram} was not conclusive because due to the scatter the 
data permit an infinity of formal solutions the goodness of which are 
statistically indistinguishably. For this reason fits were calculated, using
the Simplex algorithm (Caceci \& Cacheris \cite{Caceci}),
keeping the eccentricity fixed and permitting the
other orbital parameters to be adjusted. As illustrative cases, the dotted 
and dashed curves in Fig.~\ref{o-c-diagram} show solutions for $e=0.3$ and 
$e=0.6$, respectively. They fit the data just as well as the sine curve does.
Reminding the reader that the observational timing error of the eclipse
epochs was determined such that $\chi^2_{\rm r}$ is unity in the case of a
purely sinusoidal shape of the cyclic period variations
(Sect.~\ref{Ephemerides}), $\chi^2_{\rm r} = 0.99\, (1.01)$ for
$e=0.3 \, (0.6)$ is found. It is thus not possible to prefer one 
solution over the other on statistical grounds. Therefore, the 
orbital eccentricity of the third body is not constrained by the current
data.

\subsection{Evolutionary considerations}
\label{Evolutionary considerations}

Let us now address the question whether the existence of a planet with the
characteristics derived in the previous section is compatible with the 
evolution of the V893~Sco system from a wide binary to a cataclysmic variable.
This implies that the planet would have existed before the common envelope (CE)
phase\footnote{In order to overcome difficulties explaining what might be
the configuration of a current CV and PCEB planetary system before the CE phase
it has been speculated (e.g.\ Hessman et al.\ \cite{Hessman}, Zorotovich \&
Schreiber, \cite{Zorotovic}) that the planets could have 
formed after the CE phase from matter expelled from the primary star and 
then recaptured to form a circum-binary disk. In this case the currently
discussed question is, of 
course, irrelevant.}. The subsequent considerations are not meant to 
substitute a rigorous in-depth study. The aim is just to get a general 
idea if serious problems become obvious or not.

As criterion for the planetary scenario to be viable from the 
evolutionary point of view it is assumed that the orbit of the planet is
larger than the radius $R_{\rm CE}$ of the common 
envelope. $R_{\rm CE}$ is approximated by the radius $R_{\rm max}$ which a
single star has when its core mass is equal to the mass of the white
dwarf in V893~Sco. 
For 0.75 (0.5, 1.0) $M_{\sun}$, i.e.\ the nominal white dwarf mass adopted in
the previous section, minus and plus the assumed uncertainty, the 
core mass -- radius relation for giants as given by 
Joss et al.\ (\cite{Joss}) yields $R_{\rm max} = 2.8$\, (0.9, 4.6) AU. 

This radius cannot directly be compared to that of the orbit of the third
body as derived in the previous section because the distance between the
planet and the inner binary will have changed
during the evolution of the system. In a first approximation only 
the mass loss of the primary star during the common envelope 
phase is regarded which leads to a widening of 
the planetary orbit. Therefore, the radius of the progenitor star must
be compared to the original separation of the third body from the inner
binary. To this end one needs to know the mass of the white dwarf progenitor.
Several initial -- final mass relations for white dwarfs have been published in
the literature. For the nominal white dwarf mass of 
$M_{\cal A} = 0.75\, M_{\sun}$ the  
relation of Zhao et al. (2012) yields a mass of $M_{\rm prog} = (4.1 \pm 1.2)\, 
M_{\sun}$ for the progenitor star. This is equal to the upper limit
of progenitor star masses for which their relation is valid. The lower limit
of the validity range ($1.1 M_{\sun}$) corresponds to a white dwarf mass of 
$(0.53 \pm 0.05)\, M_{\sun}$, close to the lower limit adopted in the 
previous section. Similarly, eq. (1) of Salaris et al.\ (\cite{Salaris}) gives
$M_{\rm prog} = 3.4\, M_{\sun}$, while their alternative eq. (2) leads to
$M_{\rm prog} = 3.1\, M_{\sun}$. For the lower white dwarf 
mass limit the relations of Salaris et al.\ (\cite{Salaris}) 
have no solutions while
for the upper limit in both cases $M_{\rm prog} = 6.4\, M_{\sun}$. Finally,
using eqs.\ (2) and (3) of Catal\'an et al. (\cite{Catalan}) I find
$M_{\rm prog} = 3.2 \pm 0.2\, M_{\sun}$ 
($0.7 \pm 0.2\, M_{\sun}$;
$5.0 \pm 0.3\, M_{\sun}$) for
$M_{\cal A} = 0.75\, M_{\sun}$ 
($0.5\, M_{\sun}$; $ 1.0\, M_{\sun}$). 
To be specific and 
since I am only interested in an order of magnitude estimate, I adopt 
$M_{\rm prog} = 3.5\, M_{\sun}$ with lower and upper limits of 1.0 and 5.5 
$M_{\sun}$, respectively.

In order to calculate the radius $a_{\rm preCE}$ of the orbit of the third 
body before the
CE phase it is assumed that its orbital angular momentum $m$ is the
same before and after that phase (i.e, it remains outside the common
envelope). Knowing the mass of the planet and its current orbital radius
and period the angular momentum can be calculated which can then be used
to derive the original period $P_{\rm preCE}$ as a function of $a_{\rm preCE}$. 
Kepler's third law together with the component masses before the common 
envelope phase provided a second relation between $P_{\rm preCE}$ and 
$a_{\rm preCE}$. This enables to pin down the radius. It turns out that
$a_{\rm preCE} \propto (m/M_{\cal P})^2$. Since $m$ is also proportional to 
$M_{\cal P}$ the dependence on the planetary mass cancels out and
$a_{\rm preCE}$ does not depend on the mass and thus (via eq.~\ref{eq. 2}) on
the inclination of the outer orbit.

Assuming the mass of the secondary star not to change during the common 
envelope phase,
$a_{\rm preCE} = 1.1 \pm 0.6\, {\rm AU}$ in the case of the nominal white
dwarf mass ($0.75\, M_{\sun}$) and 
$a_{\rm preCE} = 3.7 \pm 2.0$ ($0.7 \pm 0.4$) AU in the case of the lower
(upper) limit, respectively. Here the errors do not include the
(possibly major) contribution due to the error of $M_{\rm prog}$. 

\begin{table}
\caption{Relationship between the white dwarf mass $M_{\rm WD}$ of 
V893~Sco, 
the mass $M_{\rm prog}$ of its progenitor, the maximum radius $R_{\rm max}$ of 
the progenitor and the radius $a_{\rm preCE}$ of the planet before the CE phase}
\label{Relationships}
\centering

\begin{tabular}
{l@{\hspace{4ex}}c@{\hspace{1ex}}c@{\hspace{1ex}}c@{\hspace{1ex}}}
\hline\hline
$M_{\rm WD}\, (M_{\sun})$       & 0.5  & 0.75 & 1.0 \\ 
\hline
$M_{\rm prog}\, (M_{\sun})  $   & 1.1  & 3.5  & 5.5 \\
$R_{\rm max}$ ($\approx R_{\rm CE}$)\, (AU)   & 0.9  & 2.8  & 4.6 \\
$a_{\rm preCE}\, ({\rm AU})$ & 3.7  & 1.1  & 0.7 \\
\hline
\end{tabular}
\end{table}

Summarizing, in Table~\ref{Relationships} the mass of the white dwarf
progenitor, the maximum radius that it would have as a single star (which 
is expected to be of the order of the radius of the CE) and the initial radius
of the planetary orbit for the three adopted values of the white dwarf mass
are listed, the current orbital radius of 4.5~AU of the planet having been 
used in the calculations.
Comparing $a_{\rm preCE}$ with $R_{\rm max}$ shows that in the
framework of the simple scenario explored here the third body orbit 
should have been larger than the radius of the white dwarf progenitor at 
the onset of the common envelope phase only if the primary of 
V893~Sco has a mass somewhat less than the average for CVs (but still 
within the range of primary masses found in cataclysmic variables). Thus, the
presence or not of a conflict depends decisively on the white dwarf mass.
A possible problem could be alleviated if
the assumption of angular momentum conservation is not valid. The
planet could have lost angular momentum due to friction if it revolved
within the outer reaches of the CE or a dense wind from the inner binary.
Depending on the balance of angular momentum loss and decrease in the total 
system mass this could have caused an smaller increase in the orbital radius 
than in a scenario without loss of angular momentum, or even a decrease.

A more rigorous investigation of the fate of planets considering the
evolution of the parent star through the giant phase has recently been
performed by Nordhaus \& Spiegel (\cite{Nordhaus}). Their scenario differs 
somewhat from the current one because they only regard planets around single
stars, not binaries. However, since in the above simplified calculations
the binary nature of the V893~Sco progenitor system did not enter 
explicitly a comparison should still be possible. In fact, the detailed 
study of Nordhaus \& Spiegel (\cite{Nordhaus}) does confirm the results 
obtained here. Their Fig.~3 shows that for a stellar mass equal to the 
minimum mass considered presently, a 10~$M_{\rm Jup}$ mass planet with an 
orbital radius $>$2.5~AU should escape engulfment by the parent star. For 
a $\sim$$1.5\, M_{\sun}$ star (equivalent to a white dwarf of $0.57\, M_{\sun}$ 
according to Catal\'an et al., \cite{Catalan}) this limit is $\sim$3.5~AU.

Another question regards the stability of the planetary orbit before the
common envelope phase in a situation where its radius is not
necessarily large compared to the separation of the binary star components.
This issue was investigated by Holman \& Wiegert (\cite{Holman}) for planets 
on circular orbits and coplanar with the binary orbit. In the most favourable
situation of a binary with negligible orbital eccentricity and assuming
a reduced mass ratio $\mu = M_{\cal B}/(M_{\rm prog}+M_{\cal B}) = 0.11$ 
(corresponding to the lower limit of the masses for V893~Sco) 
the planetary motion should be stable if the ratio between the semi-major 
axis of the planetary orbit is more then twice that of the binary orbit. As 
Tab.~\ref{Relationships} shows this is indeed the case if the mass of the 
white dwarf in V893~Sco is on the low side.   

I therefore conclude that these evolutionary considerations do not reveal
insurmountable obstacles to the presently regarded interpretation of the
observed cyclic period changes of V893~Sco if the white dwarf is not
significantly more massive than $\approx 0.5 - 0.6\, M_{\sun}$ as estimated 
by Matsumoto et al.\, (2000).

\section{Frequency analysis}
\label{Frequency analysis}

Warner (\cite{Warner04}) gives an overview over the rich phenomenology of
oscillations in the brightness of many cataclysmic variables. These are
generally divided into two classes termed dwarf nova oscillations (DNOs) and
quasi-periodic oscillations (QPOs).

DNOs are normally (but not exclusively) observed 
in dwarf novae during outburst. They are of low  
(some millimagnitudes) and often strongly variable amplitude 
and usually have periods in the general range of a 
couple of seconds up to about a minute. They exhibit a wide range of coherence: 
Over the course of hours the periods can change by many seconds, but
they also may remain stable on the millisecond level. 
Warner et al.\ (\cite{Warner03}) also introduce a second kind of DNOs which 
they term ``longer period DNOs (lpDNOs)'' with periods typically 4 times
longer than normal DNOs.

Quasi-periodic oscillations have longer periods than DNOs; 
of the order of several hundred seconds.
They are more elusive than DNOs because they are much less coherent. In
general, they vanish after only a handful of periods [see Warner 
(\cite{Warner04}) for a description of their phenomenology and their 
relationship to DNOs]. This makes it
difficult to distinguish them from the ubiquitous stochastic flickering 
activity in CVs. Therefore Warner suspects that {\em ``some
reclassification {\rm [between flickering and QSOs]} may be necessary; 
some of the flickering and flaring commented on in the CV literature has a 
QPO look about it''}. On the other hand, the opposite may also be true. An
accidental superposition of unrelated flickering events occurring on
similar time and magnitude scales may well mimic QPOs. I will investigate
the probability for this to happen in Sec.~\ref{Numerical experiments}.
Moreover, if individual flares in the flickering activity of CVs are not
independent but physically connected, they may well assume the 
characteristics of QSOs. The
borderline between flickering and QSOs then becomes blurred and the question
may be raised if there is a conceptual difference between these two 
phenomena at all.

Bruch et al.\ (\cite{PaperI}) were the first to claim the detection of QPOs 
with a period of 342~sec in V893~Sco in the light curve of 
1999 June 21/22. Warner et al.\ (\cite{Warner03}) find
that the star {\em ``frequently has large amplitude QSOs''}, but they give
details of only one observation on 2000 June 1, when V893~Sco
exhibited QSOs with a period of 375~sec concurrently with DNOs at 
25.2~sec which appeared only during part of the observing 
run\footnote{Warner et al.\ (\cite{Warner03}) say that these observations
were performed during outburst. However, in their Tab.~1 they quote that
V893~Sco had a mean visual magnitude out of eclipse of $14\fm1$. 
This is typical for the quiescent state of the star (see 
Fig.~\ref{long-term-lc}). An inspection of the AAVSO light curve shows 
that V893~Sco had just declined
from an outburst.}. Pretorius et al.\ (\cite{Pretorius}) did not observe QSOs
but report DNOs with a period of 41.76~sec during a 0.9 hour section of a 
light curve obtained on 2003 May 21\footnote{Pretorius et al.\ 
(\cite{Pretorius}) quote a mean magnitude of $15\fm1$ for these observations,
significantly lower than normal quiescence (see Fig.~\ref{long-term-lc}). 
AAVSO data show an outburst at 
$12^{\raisebox{.3ex}{\scriptsize m}}_{\raisebox{.6ex}{\hspace{.17em}.}}7$
about 3 days earlier, and 1.5 days later a typical quiescent magnitude of 
$14^{\rm m}$.}. 

In order to search for oscillations in the light curves of 
V893~Sco presented in this paper a frequency
analysis at high, medium and low frequencies was performed. Here, low
frequencies are understood to refer to variations on timescales 
$\tau \ge 15^{\raisebox{.3ex}{\scriptsize m}}$. This includes the orbital
period and thus the range typical for positive or negative superhumps
observed in many CVs. Positive superhumps are restricted to systems with
superoutbursts -- which are not observed in V893~Sco -- and some 
novalike systems (permanent superhumpers). They are therefore not expected 
in the present case. On the other hand, negative superhumps are sometimes
observed in different types of CVs (Wood \& Burke, \cite{Wood07}), including
dwarf novae. Medium frequencies are defined to encompass the range 
$15^{\raisebox{.3ex}{\scriptsize m}} \ge \tau \ge 2^{\raisebox{.3ex}{\scriptsize m}}$,
where QPOs are expected to be observed if they are present, 
and high frequencies refer to $\tau \le 2^{\raisebox{.3ex}{\scriptsize m}}$, the
typical range for DNOs. In all cases the eclipses were
masked before performing the analysis.

In view of the strong flickering variability which easily masks low amplitude
oscillations the light curves were pre-whitened in the following way: For the 
high frequency analysis the light curves were binned in intervals of 
$2^{\raisebox{.3ex}{\scriptsize m}}$. 
A spline was then fitted to the resulting data points and
subsequently subtracted from the original data. This procedure removes --
if not completely -- to a large degree variations on timescales above those 
defined by the data binning.
For the medium frequency analysis the data were prepared in a similar manner,
binning the light curves in intervals of $15^{\raisebox{.3ex}{\scriptsize m}}$. 
No pre-whitening was performed for the low frequency analysis.
Power spectra of light curves were then calculated using
the Lomb-Scargle algorithm (Lomb \cite{Lomb}, Scargle \cite{Scargle}, 
Horne \& Baliunas \cite{Horne}). 

\subsection{Medium frequencies and the interpretation of peaks in power spectra}
\label{Medium frequencies and the interpretation of peaks in power spectra}

I will start the discussion concentrating on medium frequencies because
this is the range where the most interesting results are observed and
where previous studies reported the detection of oscillations (\cite{PaperI},
Warner et al.\ \cite{Warner03}).
Apparently significant signals were also found in the power spectra
of other light curves of the present study.
However, before accepting them blindly as real it is
appropriate to raise the question of whether a peak in a power
spectrum may be regarded as caused by an oscillation in the
stellar light due to a real physical phenomenon or whether it is merely due
to an accidental superposition of independent random events.

As already hinted at above, it 
is not trivial to assess the significance of a peak in a power spectra 
of light curves of cataclysmic variables.
The presence of strong flickering easily causes signals to be present
at apparently stochastically distributed frequencies,
reflecting merely the typical timescales on which strong flickering flares 
occur. It is therefore not obvious how to distinguish between 
QSOs and pure flickering (if these are really distinct phenomena). 

In order to get a feeling for the probability that flickering could cause
power spectrum signals that mimic apparently significant oscillations 
some numerical experiments are first performed. It will then be tried 
to estimate the probability that the highest peak in the power spectrum of
each of the available light curves is significant or not. 

\subsubsection{Numerical experiments}
\label{Numerical experiments}

For the numerical experiments three related, but slightly different
procedures are pursued which will be referred to as
models 1 -- 3 subsequently. They consist of the frequency
analysis of light curves, the individual points of which have been 
stochastically shuffled. 

To generate the models the data points of a real light curve (after 
subtraction of variations on the timescale above 15 minutes as described
above) were
reshuffled in order to result in a new light curve which maintains the 
original data points, but in a random order. Repeating this procedure
many times and comparing the power spectra of the randomized light curves
with that of the original data should give an indication of the distribution 
of peaks caused by purely random flickering. 

However, in reshuffling the original data points one has to consider that 
they are not independent but correlated on the flickering timescale. 
Using the light curve of 2000 June 7/8, as reference this timescale 
was found to be 75 seconds, as
determined from the e-folding time of the autocorrelation function of the
original light curve. Therefore sections of 75 seconds (15 data points) 
of the light curve were shuffled, creating 10\,000 randomized versions
of the same lengths. The corresponding power spectra were then calculated.

The difference between models 1, 2 and 3 is as follows: In creating model 1,
the original light curve was cut into small segments of 75 seconds duration
each. The shuffled light curves consist of randomly concatenated segments until
the resulting data set has the same number of points as the original one,
permitting to pick the same section more than once. Thus, an individual shuffled
light curve may contain repeated sections, while other parts of the original
curve are not used. Model 2 is the same as model 1, however, without permitting
to pick the same section repeatedly. Therefore, all data points of the original
light curve are used once and only once. For model 3, 75 second sections of
the original light curve with arbitrary starting points were concatenated until
the resulting curve has the same length as the original one.

The different models yielded results which were consistent with each other.
Therefore, I only discuss the outcome of model 2, which 
is summarized in 
Tab.~\ref{Frequency analysis of shuffled light curves} and
Fig.~\ref{orig-vs-shuffled}. The figure shows in panel (a) the power 
spectrum of the original data. It has a couple of peaks in the 
frequency range below 3~mHz, but none of these is outstanding and would thus 
suggest a periodic or quasi-periodic oscillation, with the exception of the
peak close to 2.85 mHz. Most of the power spectra 
of the shuffled data are inconspicuous. However, in 2.2\% of all cases the 
highest power spectrum peak of a randomize light curve was higher than the 
highest peak corresponding to the original data [$N ({\rm max > max_{real}})$ 
in Tab.~\ref{Frequency analysis of shuffled light curves}], in the most 
extreme case by 38\% [$\Delta ({\rm max})$]. This is shown in 
panel (b) of Fig.~\ref{orig-vs-shuffled}. Indeed, on the basis of that power 
spectrum an uncritical observer might claim the presence of several discrete
periodicities in the stellar light!

\begin{table}
\caption[]{Frequency analysis of shuffled light curves.}
\label{Frequency analysis of shuffled light curves}
\centering

\begin{tabular}{l@{\hspace{2ex}}c}
\hline\hline
reference light curve            & BJD 2451702 \\
$N ({\rm max > max_{real}})$ (\%) & 2.2         \\
$\Delta ({\rm max})$ (\%)        & 38          \\
$N ({\rm contrast > 2.5})$ (\%)  & 0.04        \\
$N ({\rm contrast > 2.0})$ (\%)  & 0.7         \\
$N ({\rm contrast > 1.5})$ (\%)  & 6.7         \\
\hline
\end{tabular}
\end{table}

   \begin{figure}
   \centering
   \resizebox{\hsize}{!}{\includegraphics{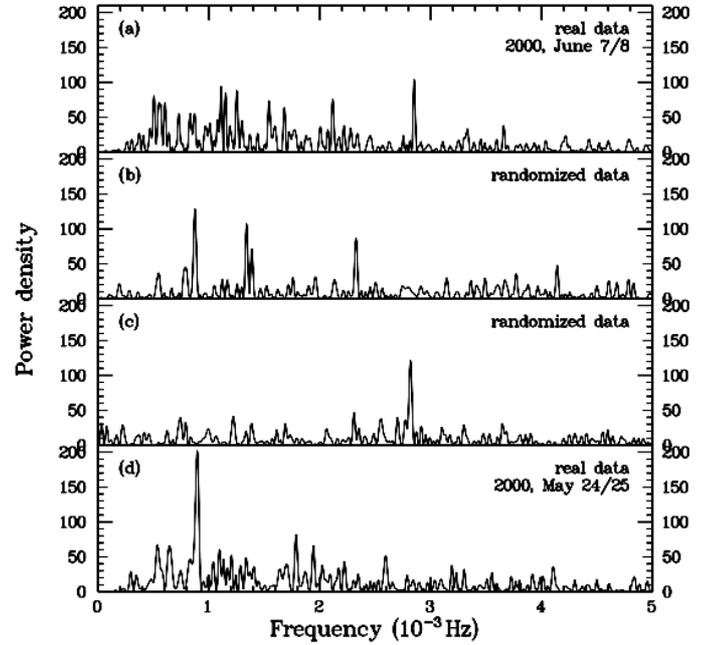}}
      \caption[]{{\em Panel (a):} Power spectrum of the original light curve of 
                 2000 June 7/9, after removal of the eclipses and variations 
                 on timescales longer than $\approx$$15^{\rm m}$.
                 {\em Panel (b):} Power spectrum of the randomized light
                 curve with the highest peak in 10\,000 trials.
                 {\em Panel (c):} Power spectrum of the randomized light
                 curve with the largest contrast between the highest and the
                 second highest peak.
                 {\em Panel (d):} Power spectrum of the original light curve of
                 2000 May 24/25. See text for details.}
\label{orig-vs-shuffled}
\end{figure}

Since the human eye is easily deceived by a 
high contrast between the dominating
power spectrum peak and the secondary peaks, suggesting an outstanding peak 
to be due to a real signal, the ratio between the highest and the second 
highest peak in the power spectra of the randomized data sets was 
also investigated. In 0.02\% (0.83\%, 8.67\%) of all cases the dominant
peak reached more than 2.5 (2.0, 1.5) times the power of the second strongest 
peak. These numbers are also listed in 
Tab.~\ref{Frequency analysis of shuffled light curves}. The
power spectrum of a randomized data set with the highest contrast between the
primary and secondary peaks is shown in panel (c) of 
Fig.~\ref{orig-vs-shuffled}, while panel (d) contains the power spectrum of 
the light curve of 2000 May 24/25 as an example of a high contrast in real 
data. The signal in the power spectrum of the randomized data 
set can easily be mistaken as real, shedding thus doubt on the reality of the 
signal observed in panel (d). Care must therefore be taken in the 
interpretation of signals as QPOs in the power spectra of flickering light
curves. This particular case will be discussed together with
similar ones in Sect.~\ref{Individual cases}.

\subsubsection{The significance of power spectrum peaks}
\label{The significance of power spectrum peaks}

For each of the observed light curves I will now try to quantify the
significance of the peaks in their power spectra. More specifically, the
following question will be addressed: What is the probability that the power 
density of a randomized light curve at any frequency within a given interval 
is higher than a certain limit. To this end the data points of the original
light curves were shuffled in a similar way as in 
Sect.~\ref{Numerical experiments}
(model 2) after determining the flickering timescale (and thus the length of
individual data segments) separately for each light curve from the width of 
the central peak of its auto-correlation function.

For each light curve 1\,000 randomized versions were created.
From the ensemble of their power spectra, for each frequency $\nu$ the 
distribution $\chi_D (\nu)$ of the 
resulting power density values $D(\nu)$ at this frequency can be constructed. 
The probability $P_\nu^- (D_0)$ to find a power density below a certain 
$D_0 (\nu)$ at the frequency $\nu$,  i.e.\ the probability for 
$D(\nu) < D_0 (\nu)$ is
\begin{equation}
P_\nu^-(D_0) = \frac{\int_0^{D_0} \chi_D (\nu)\, {\rm d}D}
{\int_0^\infty \chi_D (\nu)\, {\rm d}D} \, .
\end{equation}

Knowing thus the probability for the power density of a randomized light curve
to be lower than a given limit at a given frequency, what is the probability
$P^-(D_0)$ that the power density is lower than that limit at any 
frequency within a given interval? Provided that the frequencies are 
independent from each other this probability is simply the product of the 
respective probabilities for each frequency,
\begin{equation}
P^-(D_0) = \prod_{\nu_1}^{\nu_n} P_\nu^-(D_0) \, ,
\end{equation}
where $\nu_1 \ldots \nu_n$ are the individual frequencies.

The answer to the inverse question, i.e.\ what is the probability $P^+(D_0)$
that the power density of a randomized light curve is higher than the 
limit $D_0$ at any frequency is then
\begin{equation}
P^+(D_0) = 1 - P^-(D_0) \, .
\end{equation}
Thus, if $D_0$ is the power density at a given frequency in the power spectrum
of the real light curve -- say, the highest peak in the spectrum -- a small  
value of $P^+(D_0)$ is an indication of a true signal at that frequency.

The above reasoning only holds if the frequencies at which the power spectra
are sampled are independent. In other words, the spectra must not be 
oversampled. In order to find the appropriate step width in frequency, avoiding 
oversampling but not degrading resolution, first a highly oversampled power
spectrum was calculated for each light curve. The full width of the central
peak in its auto-correlation function was taken to be its frequency resolution
and was adopted as step width for the calculation of the power spectra of the
randomized light curves used to determine $P^+(D_0)$. 

Fig.~\ref{prob-dist} 
shows the distribution of $P^+(D_0)$ for all light curves.
$D_0$ was chosen to be the power density of the highest peak in the power
spectrum of the real data, calculated with the same step width in frequency
as in the case of the randomized light curves.

   \begin{figure}
   \centering
   \resizebox{\hsize}{!}{\includegraphics{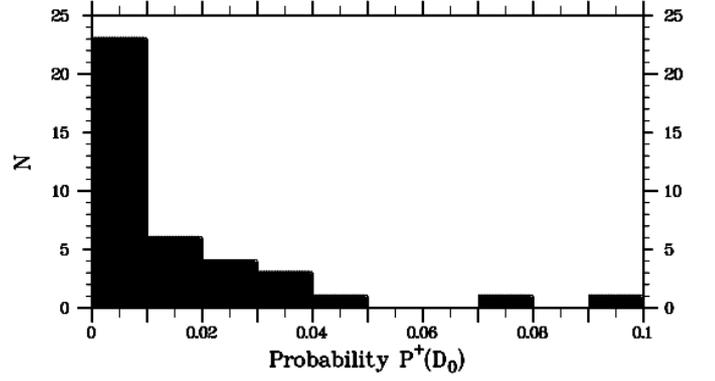}}
      \caption[]{Distribution of the probability $P^+(D_0)$ for the power
                 density of randomized light curves to be higher than the
                 highest values encountered in the power spectra of the
                 real light curves for all nightly data sets.} 
\label{prob-dist}
\end{figure}

The strong concentration of this distribution to small values of $P^+(D_0)$
suggests the conclusion that the brightness variations in the light curves 
are not completely random. To investigate this further I consider subsequently
in more detail some individual cases.

\subsubsection{Individual cases}
\label{Individual cases}

\paragraph{1999 June 21/22 (JD2451951):} 
\label{1999 June 21/22}

This is the night during which Bruch et al.~(\cite{PaperI}) 
claimed to have detected
oscillations with a period of 342~sec. The light curve is shown 
in the upper panel of Fig.~\ref{1999jun21}, while the second panel contains
the pre-whitened light curve, i.e.\ the data after removal of the eclipses 
and variations on timescales above $15^{\rm m}$ as explained above. This is 
the light curve which was actually 
submitted to the power spectrum analysis. The third panel shows the power
spectrum (normalized to the power density corresponding to the highest peak)
with the peak at 2.92 mHz which was interpreted as being due to
oscillations. In order to investigate if there is really a persistent signal 
at that frequency a stacked power spectrum was calculated in the following 
way: Sections of 1 hour duration and an overlap of 0.95 hours between
successive sections were cut out of the light curve. For each of them a
Lomb-Scargle periodogram was calculated and the individual power spectra
were stacked on top of each other to result in the two-dimensional 
representation (frequency vs.\ time) shown in the lower panel of 
Fig.~\ref{1999jun21} where the grey scale varies according to power
density. The frequency resolution is of the order of 0.4 mHz
while in temporal direction only structures separated in time by more than
1 hour are independent. 

The stacked spectra reveal that the multiple peaks in the power 
spectrum of the entire data set below $\approx$2 mHz 
can be explained by variations which persist only for
a short time and occur at seemingly random frequencies. They may therefore
readily be interpreted either as being due to short lived oscillations
that rise rapidly and die off or become incoherent after only a few 
cycles,
or by the accidental superposition of independent flares. The apparent 
oscillation at 2.92 mHz is also not
caused by a persistent signal. There is only an appreciable signal at the
beginning of the light curve (lower part of the stacked power spectrum).
Two other events at similar frequencies appear later,  
separated by time spans with no signal at all. But they are not well aligned
with the frequency observed in the third panel of Fig.~\ref{1999jun21}. It
is therefore not obvious to which degree they contribute to the peak. Even if
they do 
it is not clear whether the repeated events occurring at similar frequencies 
point at some underlying physical reason or if this is just a coincidence,
calling into question the interpretation given to the power spectrum 
in \cite{PaperI}. 

   \begin{figure}
   \centering
   \resizebox{\hsize}{!}{\includegraphics{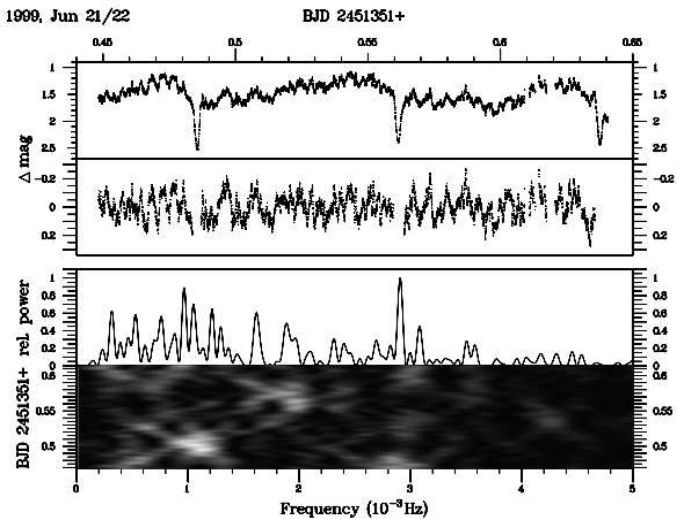}}
      \caption[]{V893~Sco during the night of 1999 June 21/22. 
                {\em Top panel:} Differential light curve (variable -- 
                 comparison star $C_1$). {\em Second panel:} Same as top panel, 
                 but after removal of eclipses and variations on timescales
                 $>$$15^{\rm m}$. {\em Third panel:} Lomb-Scargle periodogram
                 of the light curve shown in the second panel. {\em Lower
                 panel:} Stacked Lomb-Scargle periodograms of sliding sections 
                 of the light curve of panel two (see text for details).}
\label{1999jun21}
\end{figure}

\paragraph{2000 May 24/25 (JD2451689):} 
\label{2000 May 24/25} 

The results of the analysis of this night's data are shown in 
Fig.~\ref{2000m24-2} which is organized in the same way as Fig.~\ref{1999jun21}.
The power
spectrum of the pre-whitened light curve shows a strong peak at 0.89~mHz
($P = 1120^{\rm s}$). It is sufficiently outstanding to suggest that it is
really due to a periodic event in the light curve. Indeed, the results obtained
in Sect.~\ref{The significance of power spectrum peaks} place this light
curve among those for which $P^+(D_0)=0$; i.e.\ in none of the power spectra
of the 1\,000 randomized light curves investigated in that section a peak was
observed which was higher than the dominant peak corresponding to the real 
data. 

The peak at 0.89~mHz corresponds to a period of 
18$^{\raisebox{.3ex}{\scriptsize m}}_{\raisebox{.6ex}{\hspace{.17em}.}}$7
which is above the nominal cut-off of $\sim$$15^{\raisebox{.3ex}{\scriptsize m}}$
used in the pre-whitening process. In order to investigate if it is an
artefact of the data reduction the power spectrum of a light curve
pre-whitened with a cut-off of $30^{\raisebox{.3ex}{\scriptsize m}}$ was calculated.
It showed some additional features at very low frequencies, but the peak at
0.89~mHz remained unaffected, giving confidence in its reality.

The stacked power spectra show that the peak is not caused by a persistent
oscillation but rather by variations which are present during about half
of the duration of the observations. They not only vary in strength but
also in frequency close to the frequency revealed by the power spectrum of the 
entire pre-whitened light curve. Although I am reluctant to exclude that the
observed structure is due to a chance superposition of independent flares, 
variations due to a persistent but unstable physical cause appear to be more
likely in this case than during the night of 1999 June 21/22, discussed
above.

   \begin{figure}
   \centering
   \resizebox{\hsize}{!}{\includegraphics{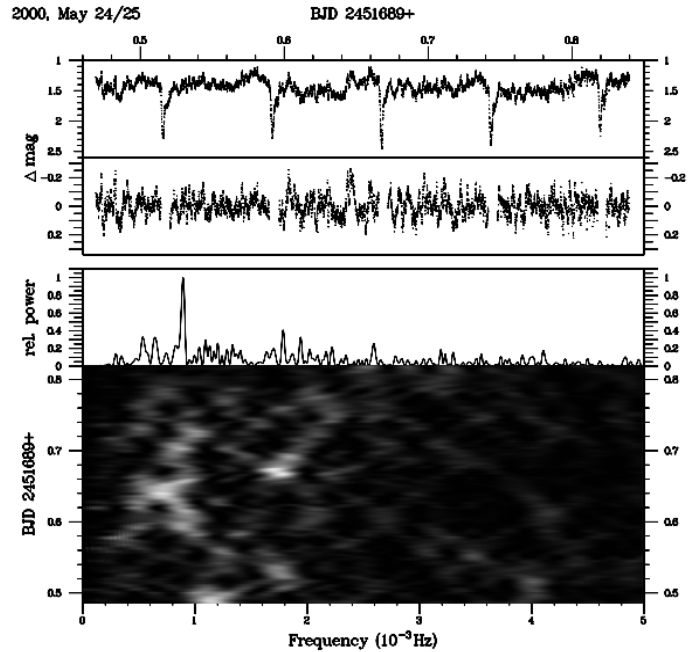}}
      \caption[]{Same as Fig.~\ref{1999jun21}, but for the night of 2000 
                 May 24/25.}
\label{2000m24-2}
\end{figure}

\paragraph{2004 July 13/14 (JD2453200):} 
\label{2004 July 13/14} 

Results for this night are shown in Fig.~\ref{2004jul13} (again, organized in
the same way as Fig.~\ref{1999jun21}). Similar to the night of 2000 May 24/25,
the power spectrum of the pre-whitened data has a strongly dominant peak at
0.86~mHz ($P=1168^{\rm s}$) which suggests some kind of persistent oscillation 
in the light curve. Also in this case $P^+(D_0)=0$ was found in
Sect.~\ref{The significance of power spectrum peaks}. The peak 
corresponding to a period longer than the nominal pre-whitening cut-off, as 
in the previous example a power spectrum calculated after a pre-whitening 
with a cut-off of $30^{\raisebox{.3ex}{\scriptsize m}}$ showed that is is not an 
artefact of the data reductions. In contrast to the other examples 
discussed in more detail in this section, V893~Sco was observed
during this night in a somewhat lower photometric state than normal quiescence, 
as mentioned in Sect.~\ref{Photometric state}. Contrary to all other
observations there is no indication of an orbital hump during this low state,
suggesting that no light modulation due to a variable visibility (or the
presence) of a bright spot occurs. Moreover, the overall variability of V893~Sco
on timescales less than $15^{\rm m}$ is higher during this night as revealed
by the increased amplitude of variations in the pre-whitened light 
curve\footnote{Note that the scale for the pre-whitened light curves is the
same in Figs.~\ref{1999jun21}, \ref{2000m24-2}, \ref{2004jul13} and
\ref{2000jun06}. This is not so for the original light curves in the 
upper panels of these figures.}.

The stacked power spectra show a signal which, while not perfectly
stable in strength over time and in frequency, persist over most of the duration
of the light curve. Of all the examples investigated so far I consider this the
most convincing of a signal which cannot easily be explained by a 
chance superposition of independent flares. It is also noteworthy that the
frequency of 0.86~mHz is close to the that observed on 2000 May 24/25 
(0.89~mHz). Moreover, apart from a particularly strong event at $\approx 1$~mHz
close to the end of the light curve (upper part of the stacked power spectrum) 
there appears to be a parallel sequence close to $\approx 1.35$~mHz at least
during a part of the observations (which 
does not show up prominently in the power spectrum of the pre-whitened light
curve). Without trying any interpretation I mention that the ratio of the two
frequencies is close to 2/3. 

   \begin{figure}
   \centering
   \resizebox{\hsize}{!}{\includegraphics{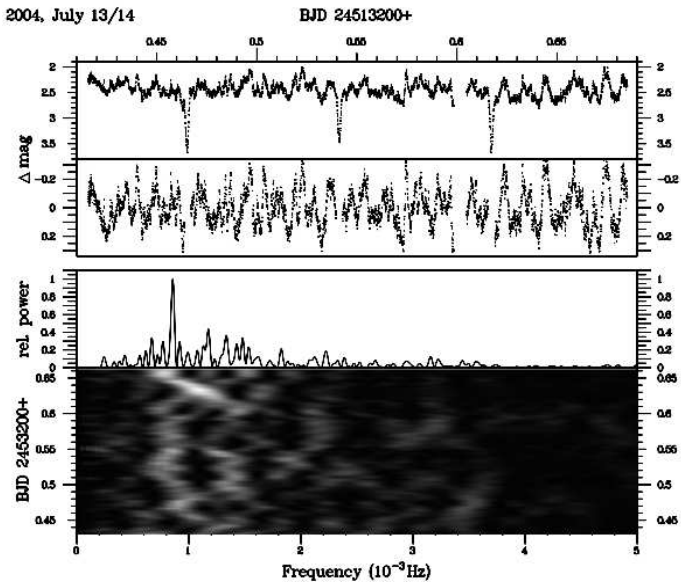}}
      \caption[]{Same as Fig.~\ref{1999jun21}, but for the night of 2004 
                 July 13/14.}
\label{2004jul13}
\end{figure}

\paragraph{2000 June 6/7 (JD2451702):} 
\label{2000 June 6/7} 

Finally, the results of the night of 2000 June 6/7 (Fig.~\ref{2000jun06}) 
are included as an example of a light curve which does not exhibit any
indication of a persistent oscillation. The power spectrum of the pre-whitened
data contain a multitude of unrelated peaks between 0.5~mHz and 2.5~mHz. But
none of them is in any way outstanding or suggests a significant signal. While
the distribution of structures in the stacked power spectra is not completely
at random\footnote{Some kind of structure must necessarily be present in the
stacked power spectra. Due to the way in which these are constructed, random
fluctuations separated in time by less than the length of the light curve
sections on which the spectra are based (1 hour in the present case) lead 
to structures connecting the discrete frequencies caused by such fluctuations.
This was verified in stacked power spectrum of light curves consisting of
pure white noise and the randomized light curves discussed in
Sect.\ \ref{Numerical experiments}.}, 
in contrast to other nights it does not suggest the presence
of anything which cannot be explained by stochastic variations in the light 
curve.

   \begin{figure}
   \centering
   \resizebox{\hsize}{!}{\includegraphics{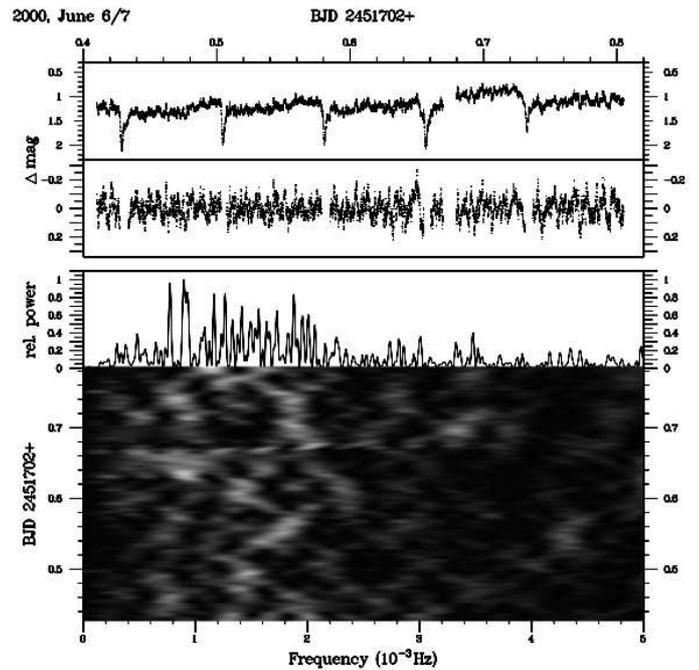}}
      \caption[]{Same as Fig.~\ref{1999jun21}, but for the night of 2000 
                 June 6/7.}
\label{2000jun06}
\end{figure}

\paragraph{{\rm This}}
small sample of individual cases should be sufficient to characterize
in a general way the occurrence of apparent or real oscillations in the
light curves of V893~Sco. The power spectra of the pre-whitened 
light curves
from the nights not discussed here all range between the extreme cases of
apparently significant signals such as on the night of 2004 Jul 13/14, or  
apparently random signals as exemplified by the data of 2000 June 6/7.
In the ensemble of data which permit to nurture the suspicion of a real 
signal no preference of a particular frequency other than the general and
wide range of 0.5 - 3 mHz can be found.

\subsection{High and low frequencies}
\label{High and low frequencies}

At high frequencies (periods shorter that $2^{\rm m}$) the power spectra of
the pre-whitened light curves did not suggest the presence of oscillations 
in any of the nightly data sets. In order to investigate if short lived 
periodicities hide in the data which do not reveal themselves in a analysis
of the complete light curves, stacked power spectra, using
sections of 900~sec duration with an overlap of 864 sec between
successive sections were calculated. 
They also did not reveal any structures which cannot
be explained by short lived random brightness fluctuations. Their amplitudes
decrease with increasing frequency as is typical for flickering
(Bruch \cite{Bruch92}). In particular, no signal which could convincingly 
be interpreted as being due to DNOs was found in the light curve of 2002
May 15/16, the only one which was observed during (the decline from) an
outburst and thus at a phase when DNOs are most often seen in other dwarf
novae (Would \& Warner, \cite{Would}). 

Not surprisingly, variations in the low frequency range (periods $>$$15^{\rm m}$)
are dominated by the orbital frequency. Analysing this range, no pre-whitening
of the light curves was performed, but the eclipses were masked. The power
spectra of all data sets of sufficient length have a strong peak close to
$f_{\rm orb} = 0.55$/hour, corresponding to the binary period, in most cases 
accompanied by a smaller peak at twice this value, i.e.\ the first overtone
of the orbital frequency. Sometimes signals at lower frequencies also
occur, caused by variations on timescales longer than the revolution period
of V893~Sco. They do not represent real periodicities as is evident from
the fact that their frequencies are not repeatable. A good 
example is shown in Fig.~\ref{lowfreq} where the upper frame contains the 
light curve of 2000 June 9/10, and the lower frame the low frequency part 
of the corresponding power spectrum. The peaks 
corresponding to the orbital frequency and its first harmonic are somewhat
offset from their nominal values, marked by dashed vertical lines. This can 
easily be explained by the strong variability of the features (hot spot and 
possibly other structure in the accretion disk) causing these signals and by
the small number of cycles covered by the light curve. The solid curve in the
upper frame is a 4-term least squares sine fit (after masking the eclipses)
to the data with periods fixed to values corresponding to the four dominating 
peaks in the power spectrum.

   \begin{figure}
   \centering
   \resizebox{\hsize}{!}{\includegraphics{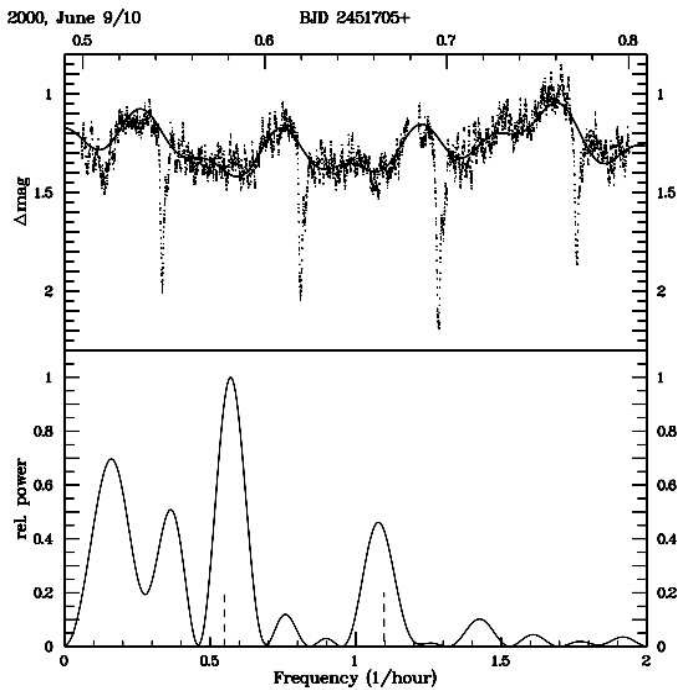}}
      \caption[]{{\em Top} Differential light curve of V893~Sco of the
                 night of 2000 June 9/10. {\em Bottom:} Low frequency
                 part of the power spectrum
                 of the light curve, after masking the eclipses. 
                 The orbital frequency and its 
                 first harmonic are marked by dashed vertical lines.
                 The solid line in the upper frame is a 4 term sine fit
                 (eclipses masked) with periods fixed to the values
                 corresponding to the dominant peaks in the power spectrum.}
\label{lowfreq}
\end{figure}
 
On some occasions light curves were observed during several subsequent 
nights (or with a gap of only one night between them). The power
spectra of the combined data sets should reveal periodic variations on
timescales significantly longer than the orbital period if they exist.
However, no consistent long-term fluctuations were found.

\section{Conclusions}
\label{Conclusions}

I have presented results of a long-term photometric study of the eclipsing
dwarf nova V893~Sco, focussing on a new determination of the eclipse
ephemerides and their interpretation, and on possible oscillations in the
brightness of the system. The main conclusions can be summarized as follows:

\begin{enumerate}

\item Normal outbursts with a comparatively low amplitude are frequent but
the absence of superoutbursts during 13 years of monitoring is unusual for
a dwarf nova with a period below the gap in the orbital period distribution
of cataclysmic variables. The likelihood for superoutbursts to hide in
the observational gaps is found to be small, unless the supercycle is close
to one year (or multiples thereof) and all superoutburst occur around
conjunction of V893~Sco with the sun.

\item The eclipse profile is strongly variable in shape, amplitude and
minimum magnitude. Large differences in the visibility of the bright spot
in the profiles of individual eclipses show that it is outshone by the 
strong ubiquitous flickering. On average, about 36\% of the total light 
of V893~Sco is occulted during mid-eclipse. 

\item Updated orbital ephemerides are calculated from eclipse timings,
leading to a significantly more precise value of the orbital period. Simple
linear ephemerides cannot adequately describe the observations. While a
quadratic term is not significant, the orbital period undergoes a cyclic
variation which leads to a sinusoidal modulation of 22.3 sec in the 
$O-C$ diagram with a period of 10.2 years.

\item These cyclic variations can be interpreted as a light travel time effect
if the presence of a giant planet with a mass of the order of 9.5 Jupiter
masses at a distance of 4.5 AU from the binary system is postulated. It was 
shown that a planet with these characteristics can survive the common
envelope phase which must have preceded the appearance of V893~Sco 
as a CV if the mass of the white dwarf in the system is on the low side of the
distribution of white dwarf masses in cataclysmic variables ($\approx 0.5 -
0.6\, M_{\sun}$).

\item A search for oscillations in the light curves of V893~Sco was
performed. At low frequencies, no persistent signal other than a modulation
with the orbital period and its first harmonic was detected. At high
frequencies a search of DNO-type oscillations met with no success. More
interesting and also more difficult to interpret are oscillations at
intermediate frequencies (timescale: $2^{\rm m} \le \tau \le 15^{\rm m}$).
In some light curves transient oscillations at frequencies between 0.5 
and 3 mHz are seen for part of the time which may be identified with QPOs.
However, an accidental superposition of unrelated flickering flares may
also lead to the observed signals. Simulations were performed in order to
gain insight to which degree random fluctuation can mimic QPOs. To this end
randomized version of real light curves were analysed. It was shown that in a
non-negligible fraction of cases their power spectra contained signals which
can be mistaken as being due to QPOs. The simple detection of an apparently 
significant signal in the power spectrum of a light curve containing strong
flickering is therefore not sufficient to claim the presence of 
QPOs\footnote{The simultaneous presence of DNOs can help (Warner,
\cite{Warner04}). Their period is
normally about 15 times less than the period of the QPOs. If DNOs at two
different frequencies are seen at the same time, their beat period corresponds
in many cases to the QPO period.}.
Even so, it is shown to be unlikely that all apparent
oscillations seen in the present ensemble of light curves can be explained
by a chance superposition of unrelated events. Therefore at least some of 
them must have a physical origin. The question to which degree QPOs and 
flickering are really conceptually different is raised.

\end{enumerate}

\begin{acknowledgements}
The observations before 2001 presented in this paper were supported 
by a grant from the Conselho Nacional de Desenvolvimento 
Cient\'{\i}fico e Tecnol\'ogico (301784/95-5).
\end{acknowledgements}


\end{document}